\documentclass[preprint,12pt]{elsarticle}

\usepackage{amssymb}
\usepackage{amsthm}
\usepackage{mathtools}

\usepackage{geometry}
\newtheorem{theorem}{Theorem}
\newtheorem{proposition}[theorem]{Proposition}
\newtheorem{lemma}[theorem]{Lemma}

\theoremstyle{definition}
\newtheorem{definition}{Definition}
\newtheorem{assumption}{Assumption}
\newtheorem*{remark}{Remark}

\usepackage{amsmath}
\usepackage{caption}
\usepackage{subcaption}
\usepackage{tikz}
\usepackage{bm}
\usepackage{float}
\usepackage{multirow}
\usepackage{colortbl}

\usepackage{tikz}
\usetikzlibrary{shapes.geometric, arrows, positioning}

\newcommand{\Nat}{{\mathbb{N}}}
\newcommand{\Real}{{\mathbb{R}}}
\newcommand{\Prob}{{\mathbb{P}}}

\newcommand{\fast}{{\textup{fast}}}
\newcommand{\slow}{{\textup{slow}}}
\newcommand{\cfast}{c^\textup{fast}}
\newcommand{\cslow}{c^\textup{slow}}
\newcommand{\tstart}{t^\textup{start}}
\newcommand{\tend}{t^\textup{end}}
\newcommand{\tpeak}{t^\textup{peak}}
\newcommand{\tbnd}{t^\textup{bnd}}
\newcommand{\tarr}{t^\textup{a}}
\newcommand{\T}{{\mathcal{T}}}

\newcommand{\X}{{\mathcal{X}}}
\newcommand{\U}{{\mathcal{U}}}
\newcommand{\A}{{\mathcal{A}}}
\newcommand{\B}{{\mathcal{B}}}
\newcommand{\Out}{{\mathcal{O}}}
\newcommand{\D}{{\mathcal{D}}}

\newcommand{\kbar}{{\bar{k}}}
\newcommand{\sfast}{{s^\fast}}
\newcommand{\sslow}{{s^\slow}}
\newcommand{\pbar}{{\bar{p}}}

\newcommand{\karma}{{\texttt{CARMA}}}
\newcommand{\nom}{{\texttt{NOM}}}
\newcommand{\toll}{{\texttt{TOLL}}}
\newcommand{\kmax}{{k_\textup{max}}}
\newcommand{\tminre}{{t^\textup{min-re}}}
\newcommand{\Are}{A_\textup{re}}
\newcommand{\bre}{{b_\textup{re}}}

\newcommand{\low}{{\textup{low}}}
\newcommand{\high}{{\textup{high}}}

\newcommand{\Ulow}{{\U_\low}}

\newcommand{\deq}{{d^{*,\textup{eq}}}}
\newcommand{\pieq}{{\pi^{*,\textup{eq}}}}
\newcommand{\deqtau}{{d^{*,\textup{eq}}_\tau}}
\newcommand{\deqlow}{{d^{*,\textup{eq}}_\low}}
\newcommand{\deqhigh}{{d^{*,\textup{eq}}_\high}}
\newcommand{\pieqlow}{{\pi^{*,\textup{eq}}_\low}}
\newcommand{\pieqhigh}{{\pi^{*,\textup{eq}}_\high}}
\newcommand{\dhom}{{d^{*,\textup{hom}}}}
\newcommand{\pihom}{{\pi^{*,\textup{hom}}}}
\newcommand{\ddelta}{{d^{*,\delta}}}
\newcommand{\pidelta}{{\pi^{*,\delta}}}

\newcommand{\pideltalow}{{\pi^{*,\delta}_\low}}
\newcommand{\pideltahigh}{{\pi^{*,\delta}_\high}}

\newcommand{\pideltatau}{{\pi^{*,\delta}_\tau}}
\newcommand{\ubar}{{\bar u}}
\newcommand{\cbar}{{\bar c}}
\newcommand{\norm}{{\textup{n}}}
\newcommand{\cbarnorm}{{\cbar^{\,\norm}}}

\newcommand{\ulow}{{u_\textup{l}}}
\newcommand{\uhigh}{{u_\textup{h}}}
\newcommand{\floor}[1]{\left\lfloor #1 \right\rfloor}
\newcommand{\ceil}[1]{\left\lceil #1 \right\rceil}

\DeclareMathOperator*{\E}{\mathbb{E}}

\newcommand{\mc}[1]{\mathcal{#1}}

\definecolor{karmagreen}{RGB}{98, 176, 69}

\usepackage[acronym]{glossaries}
\newacronym{VOT}{VOT}{Value of Time}
\newacronym{DPG}{DPG}{Dynamic Population Game}
\newacronym{DPGs}{DPGs}{Dynamic Population Games}
\newacronym{SNE}{SNE}{Stationary Nash Equilibrium}
\newacronym{HOT}{HOT}{High Occupancy Toll}
\newacronym{HOV}{HOV}{High Occupancy Vehicle}
\newacronym{MDP}{MDP}{Markov Decision Process}
\newacronym{iid}{i.i.d.}{independent and identically distributed}

\journal{Transportation Science}

\begin{document}

\begin{frontmatter}



\title{CARMA: Fair and efficient bottleneck congestion management via non-tradable karma credits}


\author[inst1]{Ezzat Elokda}
\author[inst1]{Carlo Cenedese}
\author[inst1,inst2]{Kenan Zhang}
\author[inst1]{Andrea Censi}
\author[inst1]{John Lygeros}
\author[inst1]{Emilio Frazzoli}
\author[inst1]{Florian D\"orfler}

\affiliation[inst1]{organization={ETH Zürich},
            city={Zürich},
            country={Switzerland}}

\affiliation[inst2]{organization={EPFL},
            city={Lausanne},
            country={Switzerland}}


\begin{abstract}

This paper proposes a non-monetary traffic demand management scheme, named $\karma$, as a fair solution to the morning commute congestion.
We consider heterogeneous commuters traveling through a single bottleneck that differ in both the desired arrival time and \gls{VOT}. We consider a generalized notion of \gls{VOT} by allowing it to vary dynamically on each day (e.g., according to trip purpose and urgency), rather than being a static characteristic of each individual.
In our $\karma$ scheme, the bottleneck is divided into a fast lane that is kept in free flow and a slow lane that is subject to congestion. 
We introduce a non-tradable mobility credit, named \textit{karma}, that is used by commuters to bid for access to the fast lane. Commuters who get outbid or do not participate in the $\karma$ scheme instead use the slow lane. 
At the end of each day, karma collected from the bidders is redistributed, and the process repeats day by day.
We model the collective commuter behaviors under $\karma$ as a \gls{DPG}, in which a \gls{SNE} is guaranteed to exist. Unlike existing monetary schemes, $\karma$ is demonstrated, both analytically and numerically, to achieve a) an equitable traffic assignment with respect to heterogeneous income classes and b) a strong Pareto improvement in the long-term average travel disutility 
with respect to no policy intervention. With extensive numerical analysis, we show $\karma$ is able to retain the same congestion reduction as an optimal monetary tolling scheme under uniform karma redistribution and even outperform tolling under a well-designed redistribution scheme. 
We also highlight the privacy-preserving feature of $\karma$, i.e., 
its ability to tailor to the private preferences of commuters without centrally collecting the information.

\end{abstract}




\end{frontmatter}



\section{Introduction}
\label{sec:intro}

For decades, traffic congestion has been the source of considerable social cost to major cities around the world.
According to the INRIX $2019$ Global Traffic Scorecard report, drivers in the U.S. lost an average of 99 hours a year due to congestion. The situation was even worse in the U.K., where drivers on average spent 115 hours in traffic~\cite{inrix2019global}. 
To manage rush hour traffic, a wide variety of tools have been proposed in the literature as well as implemented in practice. Among them, congestion pricing is the most widely known due to its theoretical efficiency. 
The main idea is to internalize the negative externalities into the travel cost of individual travelers~\cite{pigou1920economics,vickrey1969congestion,de2011traffic}.
However, the classical congestion pricing is arguably unfair \cite{brands2020tradable} as it tends to favor wealthier travelers~\cite{arnott1994welfare,evans1992road,taylor2010addressing}.
Moreover, it is difficult to determine and implement the theoretically optimal tolls in real time~\cite{de2011traffic,lindsey2006economists,yang2004trial}.
As a result, most existing congestion pricing schemes are restricted to zone-, cordon- and facility-based schemes with flat rates during certain time windows~\cite{de2011traffic,cenedese:2022:EJC_bn_evs}.

Due to the limitations of congestion pricing, growing attention has been devoted to alternative quantity- and credit-based approaches. The former directly limit the number of vehicles on the road, e.g., through license-plate rationing~\cite{wang2010traffic,han2010efficiency} or highway reservation~\cite{teodorovic2005highway,edara2008model}, while the latter assign a limited number of travel credits/permits to travelers which can be traded in a monetary market~\cite{verhoef1997tradeable,yang2011managing,blum2022conceptualizing} or used as vouchers to pay for monetary tolls~\cite{decorla1994applying,daganzo1995pareto,decorla2000fair,kockelman2005credit}.
These approaches are preferable to classical congestion pricing as they avoid a net financial flow from road users to authorities, as well as mitigate some of the aforementioned practical difficulties~\cite{brands2020tradable,nie2013managing}.
Nevertheless, they fail to fundamentally address the fairness issue.
Indeed, wealthy travelers are still being favored by credit-based schemes since they have a larger capacity to buy or bypass credits than others~\cite{xiao2013managing}. As a result, more transport capacity would be allocated to travelers with higher income levels. The license-plate rationing implemented in China is emblematic in that it has induced wealthy travelers to purchase additional vehicles~\cite{nie2017license}.

A fair and efficient mechanism should allocate transport capacity based on the \emph{true urgency} of the travelers that could vary on a daily basis.
Therefore, rather than quantifying \gls{VOT} as a static value determined by the income level, we model it as a dynamic process over days. To the best of our knowledge, the daily variation in \gls{VOT} has not been widely investigated in the literature.
While a clear correlation to trip distance and purpose has been observed empirically~\cite{wardman1998value,zamparini2007meta,batley2010updating,borjesson2012income}, few studies have recognized that \gls{VOT} for the same trip (e.g., daily commute) could also change over time. A notable exception is \cite{borjesson2012income}, which reports a correlation between \gls{VOT} and the required punctuality of a trip.

\begin{figure}[tb]
\centering
\includegraphics[clip,trim=0 0 0 10, width=\textwidth]{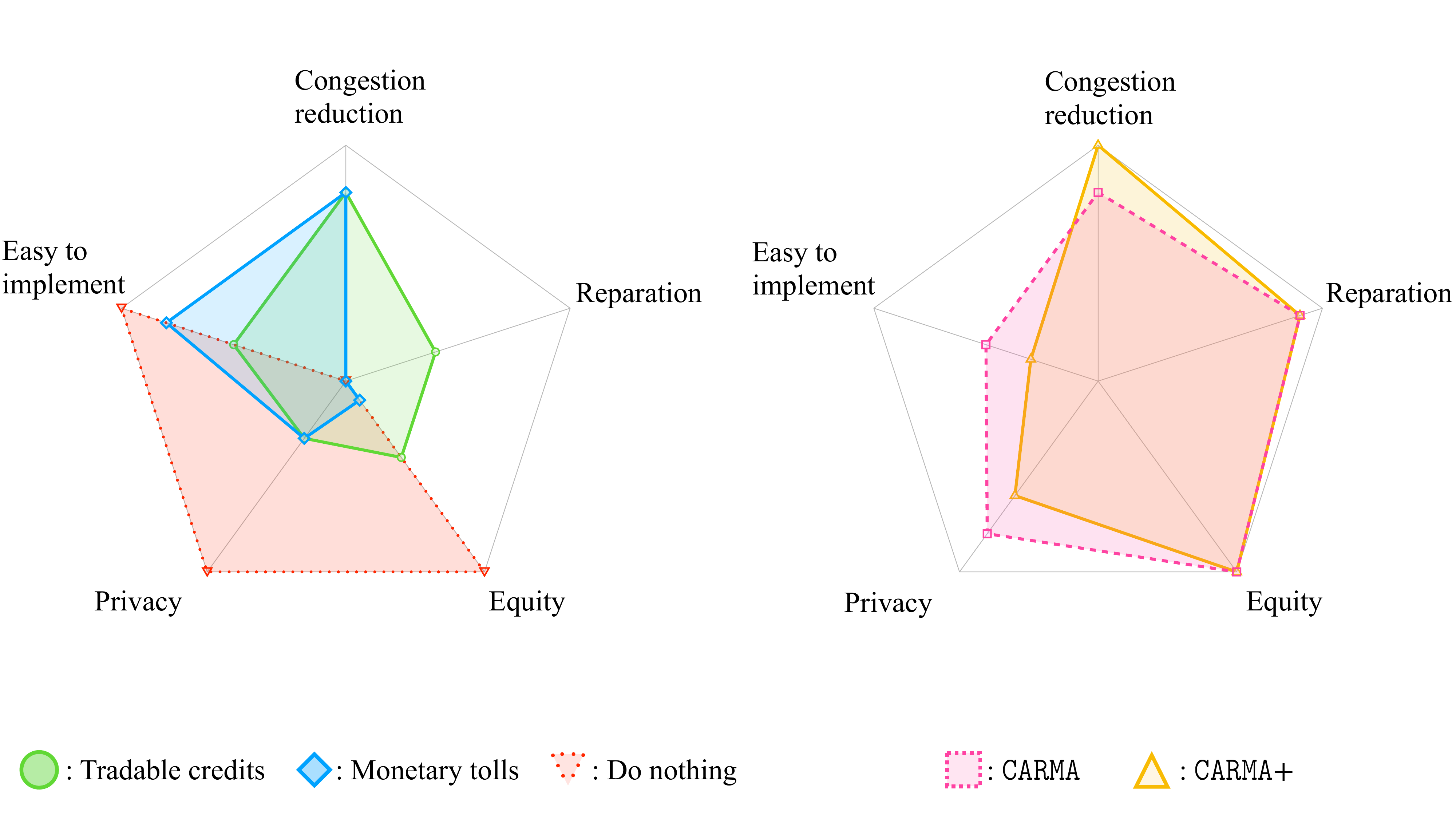}
\caption{A qualitative depiction of the performances of  existing congestion management schemes (on the left) and $\karma$ (on the right).}
\label{fig:spdergraph}
\end{figure}

In this study, we propose $\karma$, a novel congestion management scheme that uses non-monetary mobility credits called \textit{karma} to overcome the shortcomings of exisiting approaches. 
Under $\karma$, the bottleneck is divided into two lanes: a fast lane that is kept in free flow and a slow lane that is subject to congestion. 
All commuters are initially endowed with an equal amount of karma, which they use to bid for the fast lane on a daily basis. At the end of each day, the total karma collected from the bidding is redistributed to the commuters.

Figure~\ref{fig:spdergraph} presents a qualitative comparison between $\karma$ and existing policies (as well as the scenario without policy intervention) under different criteria. Specifically, $\karma$ refers to the baseline scheme with uniform karma redistribution, and $\karma$+ refers to the one with a well-designed redistribution scheme (see Section~\ref{sec:non-uniform-redistribution} for details). Similar to tradable credits, $\karma$ gives rise to a highly efficient traffic assignment, while being revenue neutral, in contrast to monetary tolls. 
When the redistribution scheme is well-designed, $\karma$ can achieve an even higher congestion reduction. 
Differently, $\karma$ manages to address the fairness issue discussed above and thus achieves higher equity among commuters with  heterogeneous income levels compared to tradable credits and monetary tolls. Moreover, $\karma$ is  more privacy-preserving, as the design of a $\karma$ mechanism does not require the private preferences of commuters.
Compared to other policies, $\karma$ compensates unfortunate commuters who fail to access the fast lane with karma that benefits them in future trips. We refer to this property as `reparation' by analogy with the commonly used term in political economy~\cite{de2008handbook,coates2015case}.
On the other hand, compared to other policies, the $\karma$ bidding mechanism is not easy to implement in real life and less intuitive for commuters, and the non-uniform karma redistribution in $\karma+$ further exacerbates such implementation issues.



The main contributions of this paper are:

\begin{itemize}
    \item We generalize the classical notion of \gls{VOT} by allowing it to vary each day according to a dynamic process (Section~\ref{sec:dynamic-VOT}).
    This is a stand-alone contribution irrespective of the proposed $\karma$ scheme, and it may be used to extend the classical results on bottleneck traffic.
    This generalization brings new insight to the interpersonal comparability of different commuters and the definition of social performance measures.

    \item We conceptualize $\karma$ and model the commuters' strategic behaviors in the framework of \acrfull{DPGs}~\cite{elokda2021dynamic}. We further prove the existence of a \acrfull{SNE} under mild conditions (Proposition~\ref{prop:sne-existence}).

    \item We prove two fundamental fairness properties of $\karma$, namely that a) it is \emph{equitable with respect to heterogeneous income} (Proposition~\ref{prop:equity}), and b) it guarantees a \emph{strong Pareto improvement} in travel costs with respect to no policy intervention under general \gls{VOT} process heterogeneity (Proposition~\ref{prop:strict-pareto}). 
    
    \item Through numerical analysis, we demonstrate that $\karma$ achieves almost the same congestion reduction compared to the optimal monetary tolling. 
    Thus, $\karma$ alleviates the classical trade-off between fairness and efficiency. To the extent of our knowledge, this is the first work that jointly satisfies strong Pareto improvement, equity with respect to income, and near-optimal efficiency.

    \item We also numerically showcase the robustness of $\karma$ towards user heterogeneity in desired arrival time. This further emphasizes the privacy-preserving feature of $\karma$, i.e., no coordination or exchange of private information (e.g., the desired arrival time) is needed to obtain an efficient equilibrium.

    \item We briefly explore how the karma redistribution scheme can be exploited to further improve efficiency and demonstrate that a better-designed mechanism, referred to as $\karma+$, could even outperform the optimal monetary tolling in congestion reduction.

\end{itemize}

We first review the related work on bottleneck congestion management in Section~\ref{sec:review}. In Section~\ref{sec:karma-bottleneck}, we detail the implementation of $\karma$, present the game-theoretic formulation and prove the existence of an equilibrium.
Section~\ref{sec:fairness} defines and analyses the key fairness properties of $\karma$. This is complemented by a comprehensive numerical analysis in Section~\ref{sec:numerical-analysis}.
Section~\ref{sec:non-uniform-redistribution} is dedicated to exploring the benefits of a non-uniform karma redistribution scheme.
Finally, Section~\ref{sec:conclusion} summarizes our conclusions and outlines directions for future work.

\section{Related work}
\label{sec:review}

As a major contributor to urban traffic congestion, the morning commute problem has been studied extensively in the literature using the stylized bottleneck model~\cite{vickrey1969congestion,hendrickson1981schedule,daganzo1985uniqueness,arnott1990economics}, where commuters between a single origin-destination pair arrive at a bottleneck dynamically and form a queue due to its limited capacity. 
It is shown that, without intervention, the selfish choice of departure time by the commuters leads to costly traffic congestion. In contrast, a time-varying monetary toll, known as Vickrey's toll, is theoretically capable of achieving the socially optimal traffic assignment where the bottleneck congestion is completely eliminated.

The original bottleneck model has been extended in various directions. One of them regards the design of a special lane that can only be used by a certain group of vehicles. The discussion started with the \gls{HOV} lanes that are exclusive for carpoolers~\cite{dahlgren1998high}, then extended to the \gls{HOT} lanes to increase the utilization rate of \gls{HOV} lanes meanwhile generating toll revenue~\cite{dahlgren2002high}. It is found that both \gls{HOV} and \gls{HOT} promote carpooling and reduce social cost, and converting an \gls{HOV} lane into an \gls{HOT} lane may further increase the carpool ratio~\cite{zhong2020dynamic}. Recent studies also discuss the capacity allocation between regular and \gls{HOV} lanes~\cite{wang2019optimal}. 
Moreover, it has been recently proposed to dedicate lanes for connected and autonomous vehicles, driven by their potential in increasing road capacity~\cite{chen2016optimal}.

Another line of research examines the performance of tradable credits in managing traffic during rush hours. 
In \cite{xiao2013managing}, Vickrey's toll is substituted by a time-varying credit charging scheme coupled with a daily initial allocation of credits. It is shown that, when late arrival is not allowed, an optimal scheme could achieve the system optimum and eliminate congestion. 
Yet, similar to Vickrey's toll, a time-varying charging scheme could be challenging to implement in practice. Hence, a step-wise charging scheme is proposed in \cite{nie2015new}. 
Specifically, commuters who pass the bottleneck during the peak time either pay a fixed amount of credits or a higher toll, while those traveling during the off-peak are rewarded with credits.
The analysis is extended in~\cite{nie2013managing} to consider elastic demand with alternative mode choice.
Although the tradable credit scheme solves several shortcomings of congestion pricing~\cite{seshadri2022congestion}, the authority still needs to design the credit charging scheme. The design is hardly optimal since the private information of commuters, e.g., their \acrshort{VOT} and desired arrival times, is often not available.
Moreover, the monetary trading of credits leads to favoring high-income travelers.

Besides tradable credits, there are other credit-based schemes proposed in the literature that do not allow the monetary trading of credits but instead use them as vouchers to pay for the monetary tolls~\cite{decorla1994applying,daganzo1995pareto,decorla2000fair,kockelman2005credit}.
In~\cite{decorla2000fair}, a lane separation scheme is proposed where the fast lane commuters  are tolled, while those on the regular lane are compensated with credits that can be used to pay for the fast-lane tolls in their future trips.
Another related work is \cite{daganzo2000pareto}, where a time-dependent toll is designed for a congestion period, but some commuters are randomly selected and exempted from paying the toll. It is proved that such policy is Pareto improving in a long time average. 
Although these two studies partially address the fairness issue, the systematic advantage of high-income commuters still persists as they have a larger capacity to pay for tolls. Consequently, commuters with different income levels do not get equal access to the fast lane.
Besides, the free access granted to commuters in \cite{daganzo2000pareto} does not align with their true need, which causes another layer of unfairness.

In essence, traffic demand management regards the efficient allocation of scarce transport resources to users with private information. 
Hence, a few studies exploited the well known auction theory to manage bottleneck traffic. 
In~\cite{wang2018trading}, the peak period is divided into small intervals and travelers may bid for each one over multiple days. To promote ride-sharing, \cite{li2022auction} designs a mechanism that jointly determines the permit allocation and the shared trip assignment. Since all of these auction mechanisms are monetary, the fairness issues still arise.

{The fairness issue regarding the monetary mechanism for scarce resource allocation is not exclusive to traffic demand management but relates to a broad array of topics such as organ donation~\cite{roth2004kidney,sonmez2020incentivized,kim2021organ} and college admission~\cite{gale1962college,hylland1979efficient}. A common approach to tackle these problems is mechanism design without money~\cite{schummer2007mechanism}.}
Despite some successes, non-monetary mechanisms are in general difficult to design in a truth-revealing manner since they cannot rely on a general incentive instrument like money~\cite{gibbard1973manipulation,satterthwaite1975strategy}.
However, a few studies have shown that truthfulness can be induced when the resource is repeatedly contested~\cite{balseiro2019multiagent}. The core principle here is to have users trade off between immediate and future access to the resource. 
In a recent study, this principle is formalized in a practical mechanism, called karma mechanism, in which users are allocated karma points to bid for the resource, and karma is transferred directly from those who manage to acquire the resource to those who yield it~\cite{elokda2023self}.
This mechanism has been recently explored in the context of static traffic routing with a centralized pricing scheme \cite{salazar2021urgency}.
In the present study, we consider dynamic bottleneck congestion without centralized control.

%

\section{CARMA: bottleneck congestion management with karma}
\label{sec:karma-bottleneck}

This section is devoted to a comprehensive description of $\karma$ and its formulation as a \acrfull{DPG}~\cite{elokda2021dynamic}. The model is built on previous studies on the bottleneck model~\cite{vickrey1969congestion,arnott1988schedule} and karma economies~\cite{elokda2023self}. A thorough review of the model preliminaries is included in Appendices \ref{sec:bottleneck} and \ref{sec:karma-preliminaries}. 
In what follows, we first define notations used in the model (Section~\ref{sec:notation}), describe how $\karma$ works in practice (Section~\ref{sec:karma-description}), introduce the notion of dynamic \acrfull{VOT} (Section~\ref{sec:dynamic-VOT}), and then specify the key components of $\karma$ in the framework of \gls{DPG}s, i.e., the immediate reward and state transition functions (Section~\ref{sec:carma-DPG}).
Finally, we define the equilibrium of $\karma$ and prove its existence (Section~\ref{sec:sne}).


\subsection{Notation}
\label{sec:notation}
Let $f : D \times C \rightarrow \Real$ be a function whose domain is defined on $D\subseteq \Nat$ and $C\subseteq \Real^n$. To distinguish the discrete and continuous arguments, we introduce the notation $f[d](c)$ in this paper, where $d\in D$ and $c\in C$.
Let $p \in \Delta(D):=\left\{\left. \sigma \in \Real_+^{|D|} \right\rvert \sum_{d \in D} \sigma[d] = 1 \right\}$ denote a probability distribution over the elements of $D$. Then $p[d]$ gives the probability of $d\in D$. Accordingly, $p[a \mid d]$ denotes the conditional probability of $a\in D$ given $d$ and $p[a \mid d](c)$ denotes the conditional probability of $a\in D$ given both $c$ and $d$. Specifically, we use $p[d^+ \mid d](c)$ to denote the one-step transition probability for $d$, which will be discussed in Section~\ref{sec:karma-transition}. 


\subsection{Practical implementation of $\karma$}
\label{sec:karma-description}
Consider a classical morning commute problem, where a population of $N$ commuters travel through a single bottleneck on a daily basis. The bottleneck with capacity $s$ [veh/min] is divided into two lanes: a fast lane with capacity $\sfast$ that is kept free of congestion and a regular (or slow) lane with capacity $\sslow$ that might be subject to congestion. Accordingly, we have $s=\sfast+\sslow$. 
To simplify the analysis, the departure window is discretized into $T$ intervals with identical length $\Delta$ and commuters are assumed to enter the bottleneck at the beginning of each interval. 

\begin{figure}[htb]
\centering
\includegraphics[width=\textwidth]{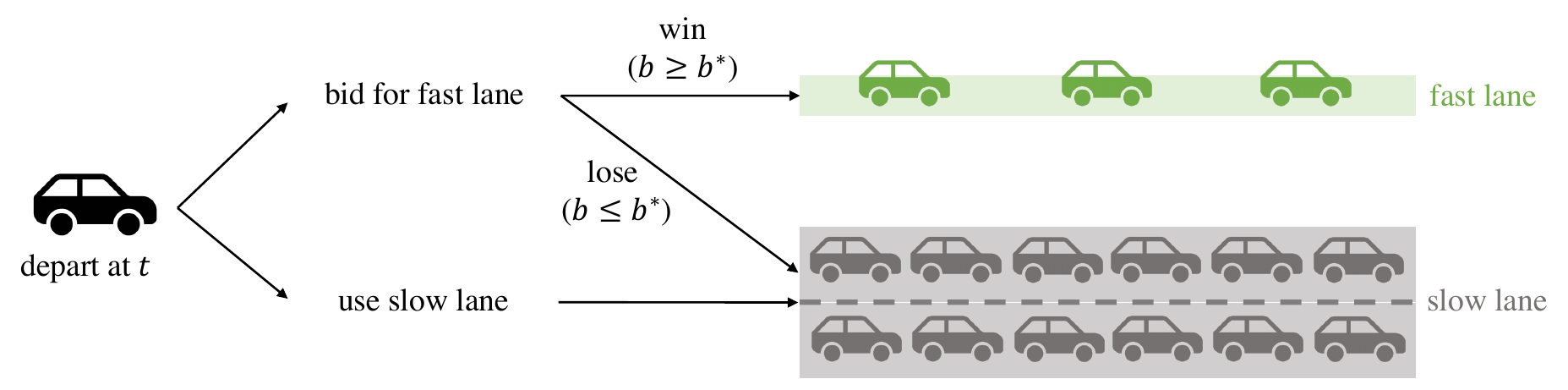}
\caption{Illustration of karma mechanism in bottleneck.}
\label{fig:karma_bottleneck}
\end{figure}

At the very beginning, each commuter is endowed with a positive integer number of non-tradable credits called karma, which can be used to bid for access to the fast lane during each time interval. Specifically, a commuter with $k\in\Nat$ karma who wishes to enter the fast lane at time $t\in\T= \{t^1,t^1 + \Delta, \dots, t^T\}$ where $t^\ell := t^1 + (\ell-1) \: \Delta$, can submit an integer bid $b\in\B[k]=\{0,\dots,k\}$ (hereafter referred to as a \emph{karma bid}). To ensure the fast lane is free of congestion, only a limited number of commuters can enter it during each interval. Hence, for each time interval $t$, there will exist a threshold karma bid $b^*[t]$. Commuters bidding above $b^*[t]$ enter the fast lane for sure, those bidding below $b^*[t]$ enter the slow lane for sure, while a random draw on the remaining fast lane capacity is used to determine who among those bidding exactly $b^*[t]$ enter the fast lane. 
Figure~\ref{fig:karma_bottleneck} gives a simple illustration of this procedure. 
The bidding process repeats for each time interval and at the end of each day, karma collected from the bidding is redistributed to commuters.

\begin{remark}[Non-$\karma$ adopters]
    For ease of presentation, we assume all commuters participate in $\karma$, though this may not be true in reality. The model can be readily modified by considering the non-$\karma$ adopters as a type of commuters who have zero karma and always choose the slow lane. The karma redistribution rule can also be easily adjusted to avoid allocating karma to them. Accordingly, their strategy set reduces to the feasible departure times as in the classical bottleneck model. 
\end{remark}

\subsection{Dynamic \acrfull{VOT} and interpersonal comparisons}
\label{sec:dynamic-VOT}

The commuters' sensitivity to time delays is modelled by the \emph{\acrfull{VOT}}, which is a conversion factor from time delay to personal inconvenience or disutility.
We consider that the \gls{VOT} is heterogeneous across both commuters \emph{and} time.
Namely, we classify commuters into a finite number of types $\tau \in \Gamma \subset \Nat$.
Commuters of the same type $\tau$ have the same desired arrival time $t^*_\tau \in \T$, set of \emph{\gls{VOT} states} $u \in \U_\tau = \{u_1,\dots,u_{U_\tau}\} \subset \Real_{>0}$, as well as \emph{\gls{VOT} process} $\phi_\tau[u^+ \mid u]$.
The \gls{VOT} process is an exogenous, irreducible Markov chain that is independent per commuter. Importantly, this modification generalizes the constant \gls{VOT} that has been commonly assumed in the literature and thus enables us to model time-varying factors in the repetitive daily commute. Specifically, commuters could be flexible with their schedules on some days (low \gls{VOT} state) but must be punctual on other days (high \gls{VOT} state).

Each \gls{VOT} process $\phi_\tau[u^+ \mid u]$ induces a stationary distribution of \gls{VOT} states $\Prob_\tau[u]$ and an associated \emph{long-term average \gls{VOT}} $\ubar_\tau = \sum_{u \in \U_\tau} \Prob_\tau[u] \: u$.
We consider that the absolute value of the \gls{VOT} state $u \in \U_\tau$ is only meaningful to individuals of the same type $\tau$ but is incomparable across types,
since each individual has its own personal scale on which to measure disutility~\cite{roberts1980interpersonal}.
To make interpersonal comparisons between commuters of different types we will consider the \emph{normalized \gls{VOT}} $u / \ubar_\tau$ rather than the absolute \gls{VOT} $u$,
see Section~\ref{sec:fairness}.

\subsection{$\karma$ as a \acrfull{DPG}}
\label{sec:carma-DPG}
A \gls{DPG}, first introduced in \cite{elokda2021dynamic}, describes the interactions among a large population of strategic agents over a dynamic process. The decision-making process of each agent is described as a \gls{MDP}, 
where the state transitions and rewards are functions of the \emph{social state} that is determined by the collective behaviors of all agents. 

\begin{figure}[htb]
\centering
\includegraphics[width=\textwidth]{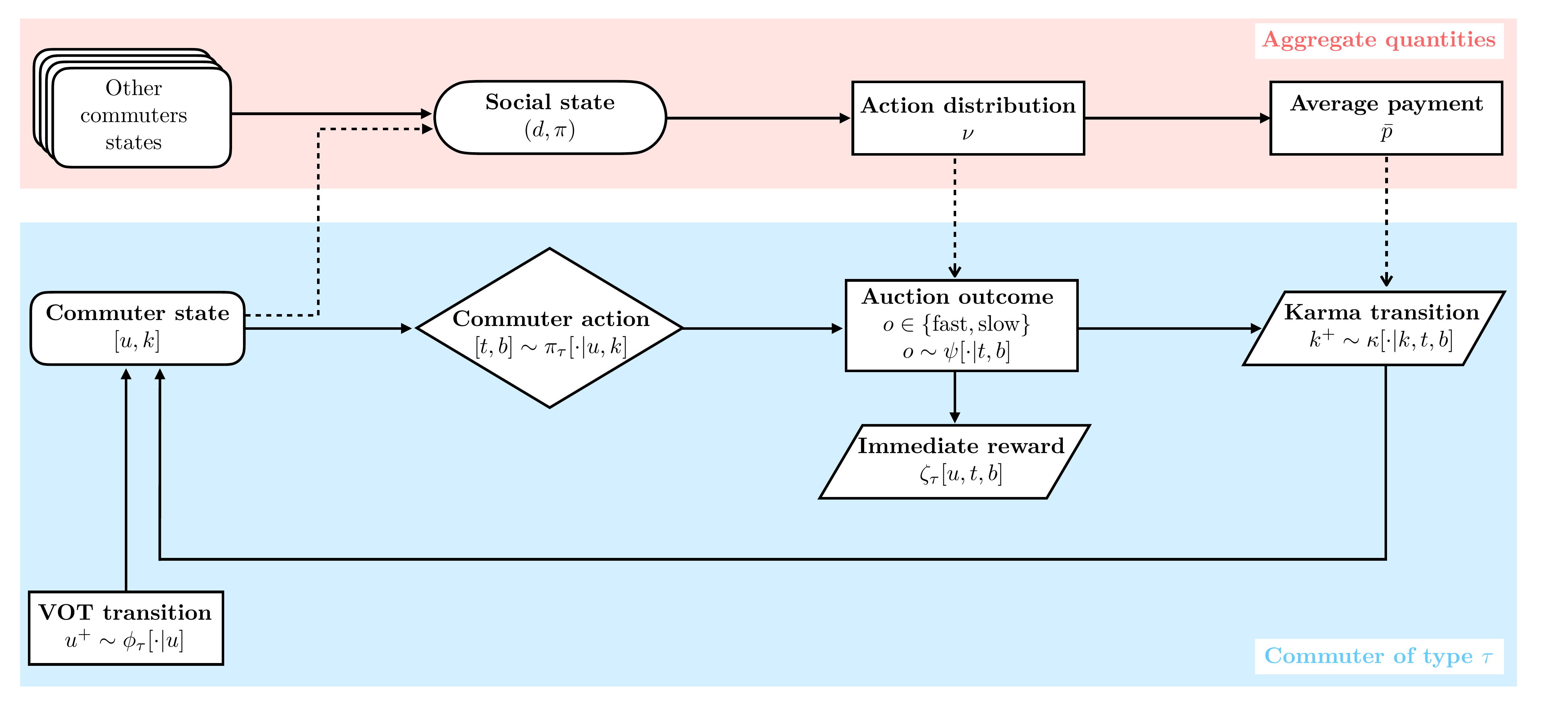}
\caption{Flowchart of a commuter of type $\tau$.}
\label{fig:flowchart}
\end{figure}

$\karma$ fits into this framework.  {Each day is modeled} as one step in the \gls{MDP} during which
the individual state is defined by the commuter's current \gls{VOT} and karma balance, denoted by the pair $[u,k]$. 
{Assuming that the goal of each commuter is to minimize the total travel cost over time}, we define the immediate reward as the negative generalized travel cost on each day in Section~\ref{sec:immediate-reward}. 
The state transition then regards the change in \gls{VOT} and karma balance between two consecutive days. The former is determined by the exogenous \gls{VOT} process while the latter depends on the karma bid distribution and the karma redistribution realized on each day,  presented in Section~\ref{sec:karma-transition}.
{Figure~\ref{fig:flowchart} schematizes the interdependence among the variables and functions used in one step of the \gls{MDP} formalized in the following sections. Specifically, it models a step of the decision process of a commuter of type $\tau$.}

\subsubsection{Immediate reward function}
\label{sec:immediate-reward}

The immediate reward function in $\karma$, denoted by $\zeta_\tau[u,t,b](d,\pi)$, gives the reward received by a commuter in type $\tau$ at \gls{VOT} $u$ who chooses action $[t,b] \in \A[k]=\T \times \B[k]$ with $t$ denoting the departure time and $b$ the karma bid for the fast lane. It is also a function of the social state $(d,\pi)$, where $d$ refers to the state distribution and $\pi$ gives the policy among all commuters (see Appendix~\ref{sec:karma-preliminaries} for details). 
Following the classic bottleneck model, we define the immediate reward as the negative total travel delay times VOT. 
The travel delay comprises two parts: queuing delay $t^q$, and early (late) schedule delay $t^e_\tau$ ($t^l_\tau$).
Let $\psi[o\mid t,b](d,\pi)$ denote the probability of a commuter entering lane $o\in\mc O=\{\textup{fast},\textup{slow}\}$ given its own decision and the social state.
Then, the immediate reward can be written as
\begin{equation}
    \label{eq:bottleneck-immediate-reward}
    \zeta_\tau[u,t,b](d,\pi)
    \coloneqq - u \: \sum_{o\in\mc O} \psi[o \mid t,b](d,\pi) \left(\alpha \: t^q + \beta \: t^e_\tau+ \gamma \: t^l_\tau\right),
\end{equation}
where $\alpha$, $\beta$, $\gamma$ are the normalized queuing delay, early arrival, and late arrival penalties, respectively. 

To ease the notation, we have omitted the dependencies of delay functions on lane assignment, departure time and social state. The full specification is given below. 
\begin{subequations}
\begin{align}
    t^q[t,o](d,\pi) &\coloneqq \begin{cases}
        0, & o=\textup{fast}, \\
        \frac{q[t](d,\pi)}{\sslow}, & o=\textup{slow},
    \end{cases} \label{eq:queue-delay}\\
    t^e_\tau[t,o](d,\pi) &\coloneqq \max\{0, \; t^*_\tau - t - t^q[t,o](d,\pi)\}, \\
    t^l_\tau[t,o](d,\pi) &\coloneqq \max\{0, \; t + t^q[t,o](d,\pi) - t^*_\tau\}.
\end{align}
\end{subequations}
In \eqref{eq:queue-delay}, $q[t](d,\pi)$ captures the queue length on the slow lane at time $t$ (i.e., the discrete equivalent of $q(t)$ defined in Appendix~\ref{sec:bottleneck}), and $t^*_\tau$ is the desired arrival time for commuters in type $\tau$.

Note that
\eqref{eq:bottleneck-immediate-reward} implies that all commuters share the same penalty ratios $\alpha/\beta$ and $\alpha/\gamma$, while the \gls{VOT} serves as a scaling factor that varies among commuters. This assumption is widely used in the literature (e.g., \cite{vickrey1973pricing} and \cite{xiao2013managing}) and formally stated below. 

\begin{assumption}
\label{as:constant_ratio_penalties}
The ratios of queuing penalty over early and late arrival penalties, i.e., $\alpha/\beta$ and $\alpha/\gamma$ respectively, are the same for all times and commuter types $\tau\in \Gamma$. 
Moreover, the normalized penalty values satisfy $\beta<\alpha<\gamma$. 
\end{assumption}

To derive the lane assignment function $\psi[o\mid t,b](d,\pi)$, we first formally define the threshold bid $b^*[t](d,\pi)$ such that 
\begin{itemize}
    \item if $b>b^*[t](d,\pi)$, then the commuter enters the fast lane, i.e., $o=\textup{fast}$;
    \item if $b<b^*[t](d,\pi)$, then the commuter enters the slow lane, i.e., $o=\textup{slow}$; 
     \item if $b=b^*[t](d,\pi)$, then the commuter enters the fast lane with a uniform probability determined by the remaining capacity and the number of commuters bidding for $b^*[t](d,\pi)$.
\end{itemize}
Let $\nu[t,b](d,\pi)$ be the mass of commuters (the share of the population) departing at $t$ and bidding $b$, then
\begin{align}
    \nu[t,b](d,\pi) = \sum_{\tau\in\Gamma}\sum_{u\in \mc U_\tau}\sum_{k\in\Nat} d_\tau[u,k] \: \pi_\tau[t,b \mid u,k],
\end{align}
where $d_\tau[u,k]$ represents the probability that a commuter is in type $\tau$ and state $[u,k]$, and $\pi_\tau[t,b \mid u,k]$ is the probability that it selects the action $[t,b]$ given state $[u,k]$. 
Then, $b^*[t](d,\pi)$ is given by 
\begin{align}\label{eq:b_star}
    b^*[t](d,\pi) = \max \left\{b\in\Nat \left| \: \sum_{b'\geq b} \nu [t,b'](d,\pi) \right.\geq \frac{\sfast}{N}\right\}.
\end{align}
Accordingly, the probability of entering the fast lane is derived as 
\begin{align}
    \label{eq:prob-outcome}
    \psi[o = \textup{fast} \mid t,b](d,\pi) = \begin{cases}
    1, & b>b^*[t](d,\pi),\\
    0, & b<b^*[t](d,\pi),\\
    \frac{\sfast/N - \sum_{b'>b}\nu[t,b'](d,\pi)}{\nu[t,b](d,\pi)}, & b=b^*[t](d,\pi),
    \end{cases}
\end{align}
and the probability of entering the slow lane is $\psi[o = \textup{slow} \mid t,b](d,\pi) = 1 - \psi[o = \textup{fast} \mid t,b](d,\pi)$.

\subsubsection{Karma transition function}
\label{sec:karma-transition}

The karma transition function encodes the rules of how karma is transferred between commuters. It depends not only on how karma is collected in the bidding process but also on how karma is redistributed to commuters. Both processes have a significant degree of design freedom. In this section, we present a simple set of rules, while more discussion on karma redistribution is included in Section~\ref{sec:non-uniform-redistribution}.

We consider that all commuters entering the fast lane pay their bids, and at the end of each day, the total payments are uniformly redistributed to all commuters, including those who entered the fast lane.
Let $p[b,o]$ be the karma payment made by a commuter who bids $b$ and enters lane $o$, then we have 
\begin{align}
    \label{eq:payment}
    p[b,o] = \begin{cases}
        b, & o=\textup{fast}, \\
        0, & o=\textup{slow}.
    \end{cases}
\end{align}
Accordingly, we may compute the \emph{average payment} $\pbar(d,\pi)$ as follows:
\begin{equation}
\begin{split}
    \pbar(d,\pi) &= \sum_{t\in \mc T}\sum_{b\in\Nat} \nu[t,b](d,\pi) \sum_{o\in\mc O} \psi[o \mid t,b](d,\pi) \: p[b,o] \\
    &= \sum_{t\in \mc T}\sum_{b\in \Nat} \nu[t,b](d,\pi) \: \psi[o=\text{fast} \mid t,b](d,\pi) \: b. \label{eq:average-payment}
\end{split}
\end{equation}
To preserve the integer value of karma, $\ceil{\pbar(d,\pi)}$ karma is distributed to a randomly selected group of commuters, and the others receive $\floor{\pbar(d,\pi)}$ karma. 
Specifically, the mass of commuters receiving $\ceil{\pbar(d,\pi)}$ karma is given by $f(d,\pi) = \pbar(d,\pi) - \floor{\pbar(d,\pi)}$.
This yields the following karma transition probabilities conditional on the outcome $o\in\Out$,
\begin{align}
    \label{eq:bottleneck-karma-tranistions-o}
    \Prob[k^+ \mid k,t,b,o](d,\pi) = \begin{cases}
        f(d,\pi), & o = \textup{fast}, k^+ = k - b + \ceil{\pbar(d,\pi)}, \\
        1-f(d,\pi), & o =  \textup{fast}, k^+ = k - b + \floor{\pbar(d,\pi)}, \\
        f(d,\pi), & o =  \textup{slow}, k^+ = k + \ceil{\pbar(d,\pi)}, \\
        1-f(d,\pi), & o =  \textup{slow}, k^+ = k + \floor{\pbar(d,\pi)}, \\
        0, &\text{otherwise}.
    \end{cases}
\end{align}
Note that \eqref{eq:bottleneck-karma-tranistions-o} does not depend on the departure time $t$ here, but we include it in the notation as we will consider time-dependent karma redistribution in Section~\ref{sec:non-uniform-redistribution}.

Finally, the karma transition function is given by 
\begin{align}
    \label{eq:bottleneck-karma-transitions}
    \kappa[k^+ \mid k,t,b](d,\pi) = \sum_{o\in\mc O} \psi[o \mid t,b](d,\pi) \: \Prob[k^+ \mid k,t,b,o](d,\pi).
\end{align}

\subsection{{Existence of $\karma$ equilibrium}}
\label{sec:sne}

The equilibrium concept for a general \gls{DPG} is the \acrfull{SNE} (see Definition~\ref{def:SNE} in Appendix~\ref{sec:karma-preliminaries}).
In brief, it describes a time-invariant social state $(d^*,\pi^*)$, which can be seen as a stochastic counterpart of the well-known Nash equilibrium. At the \gls{SNE}, no commuter has an incentive to unilaterally deviate from the policy $\pi^*$, meanwhile the state distribution $d^*$ is stationary. 

The existence of an \gls{SNE} for a general \gls{DPG} with karma requires two conditions, namely, the preservation of the total amount of karma and the continuity of functions of the social state $(d,\pi)$ (see Appendix~\ref{sec:karma-preliminaries} for details). The former can be easily satisfied by well designing the karma redistribution rule, e.g., as in \eqref{eq:bottleneck-karma-tranistions-o}, whereas the latter requires more justification.
Note that in \eqref{eq:prob-outcome}, the function $\psi[o = \textup{fast} \mid t,b](d,\pi)$ may have a discontinuity if $\nu[t,b](d,\pi)=0$, which means there is no other commuter taking the same action $[t,b]$ as the ego commuter.
We introduce the following approximation to overcome this discontinuity issue:
\begin{equation}
    \label{eq:prob-outcome-cont}
    \psi^\epsilon[o = \textup{fast} \mid t,b](d,\pi) 
    = \begin{cases}
    1, & \sum_{b'> b} \nu[b']\leq \frac{\sfast}{N} -  \nu[b] -\epsilon, \\
    0, & \sum_{b'> b} \nu[b]\geq \frac{\sfast}{N}, \\
    \frac{\sfast/N-\sum_{b'>b}\nu[b']}{\nu[b]+\epsilon}, & \text{otherwise},
    \end{cases}
\end{equation}
where $\epsilon>0$ is an (arbitrarily small) approximation parameter, and we drop the dependency of $\nu$ on $t$ and $(d,\pi)$ to ease the notation.

\begin{figure}[htb]
\begin{subfigure}{0.5\textwidth}
\centering
\includegraphics[width=\textwidth]{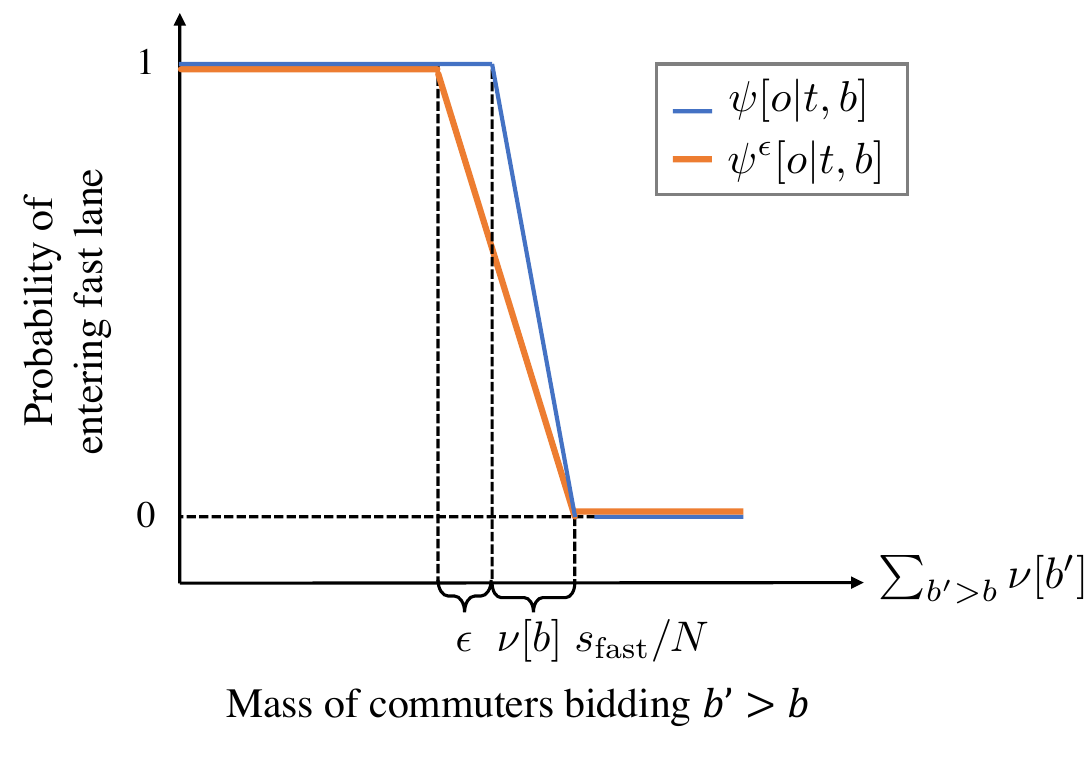}
\caption{$\nu[t,b](d,\pi)>0$}
\end{subfigure}
\begin{subfigure}{0.5\textwidth}
\centering
\includegraphics[width=\textwidth]{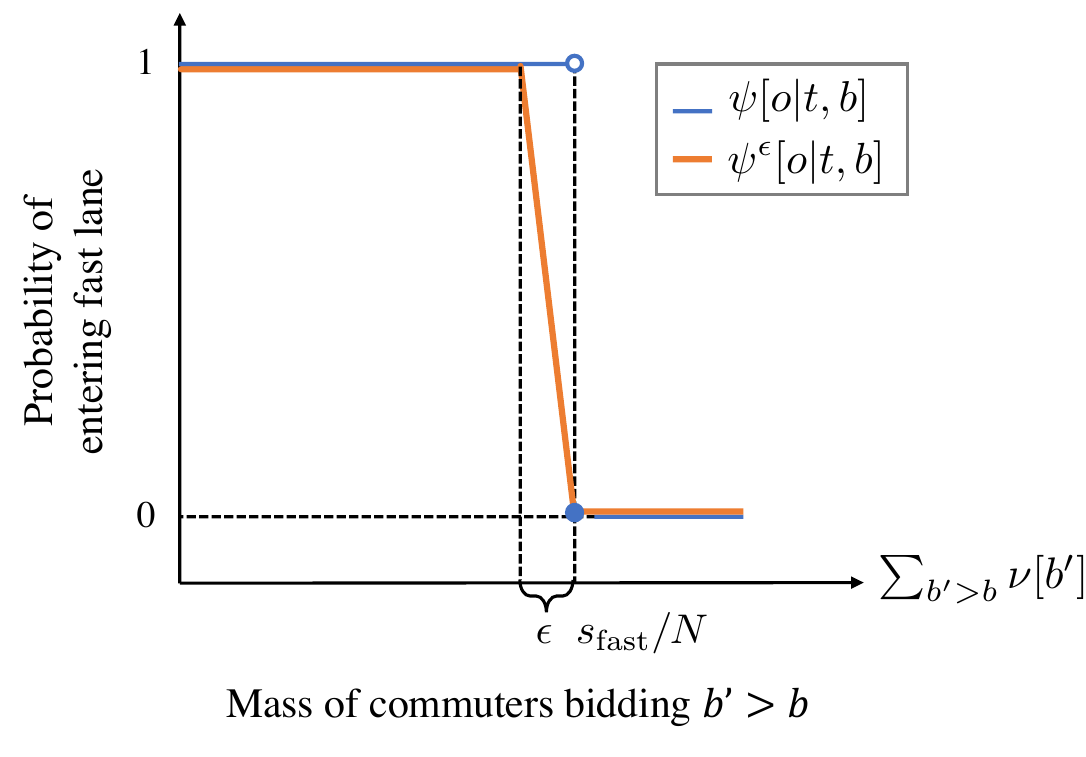}
\caption{$\nu[t,b](d,\pi)=0$}
\end{subfigure}
\caption{Illustration of the exact and approximate outcome probability.}
\label{fig:prob-outcome}
\end{figure}

A graphic illustration of $\psi^\epsilon[o  \mid t,b](d,\pi)$ and $\psi[o  \mid t,b](d,\pi)$ is shown in Figure~\ref{fig:prob-outcome}. From this and \eqref{eq:prob-outcome-cont}, we conclude that   $\psi^\epsilon$ is continuous in 
 $(d,\pi)$, and that $\psi^\epsilon \rightarrow \psi$ as $\epsilon \rightarrow 0$.
 
\begin{remark}[Approximation of $\psi$]
{The approximation parameter $\epsilon$ leads to a fast lane slightly below capacity. As per \eqref{eq:prob-outcome-cont}, commuters with the threshold bid $b^*[t]$ now have a lower probability to enter the fast lane. Consequently, only $\sfast - \frac{\epsilon}{\nu[b^*] + \epsilon}(\sfast - \sum_{b'>b^*}\nu[b^*])$ commuters enter the fast lane.}
Nevertheless, using $\psi^\epsilon[o \mid t,b](d,\pi)$ instead of $\psi[o \mid t,b](d,\pi)$ should give a negligible effect on the $\karma$ allocation as $\epsilon$ can be made arbitrarily small. 
\end{remark}

We are now ready to present the first main result of this work, that is, the existence of \gls{SNE} for $\karma$.
\begin{proposition}[Existence of a \gls{SNE}]
\label{prop:sne-existence}
There always exists a \gls{SNE} for the \gls{DPG} specified for $\karma$, which has the immediate reward function defined in \eqref{eq:bottleneck-immediate-reward} and the karma transition function defined in \eqref{eq:bottleneck-karma-transitions} with $\psi$ replaced by $\psi^\epsilon$ as per \eqref{eq:prob-outcome-cont} for some $\epsilon>0$.
\end{proposition}

\begin{proof}
The proof is included in Appendix~\ref{proof:karma-preservation}.
\end{proof}

We refer to such an equilibrium as \emph{$\karma$ \gls{SNE}} hereafter. A $\karma$ \gls{SNE} can be computed using an algorithm inspired from  evolutionary dynamics~\cite{sandholm2010population}, as described in detail in \cite[Alg.~1]{elokda2023self}.

\section{Fairness properties of $\karma$}
\label{sec:fairness}
In this section, we proceed to analyze the fairness properties of $\karma$ by investigating the normalized long-term average travel cost of heterogeneous commuters at the $\karma$ \gls{SNE}.
Specifically, we consider two kinds of heterogeneity in the \gls{VOT} processes:
\begin{itemize}
    \item[\S~\ref{sec:equity}:] Heterogeneity in \gls{VOT} scale, where commuters have the same VOT process up to a difference in scale of the VOT values.    
   We prove that $\karma$ is scale-invariant, which can be interpreted as \emph{equity with respect to income level}.
    
    \item[\S~\ref{sec:pareto-improvement}:] General heterogeneity in \gls{VOT} process. We prove that $\karma$ guarantees a \emph{strong Pareto improvement} in the normalized long-term average travel costs with respect to no policy intervention. 
\end{itemize}
In both cases, we assume that commuters share the same desired arrival time. Furthermore,  in the latter, we also introduce a mild assumption on the time discretization. These assumptions are formally presented below.

\begin{assumption}(Homogeneous $t^*$)
    \label{as:same-t*}
    All commuter types $\tau\in\Gamma$ have the same desired arrival time $t^*_\tau=t^*$.
\end{assumption}

\begin{assumption}(Time discretization)
    \label{as:discretization}
    Let $\tstart$ ($\tend$) be the start (end) time of the bottleneck queue under no policy intervention, given in~\eqref{eq:t-start-t-end-t-peak}, then the
    time is discretized such that $\tstart, \tend \in \T$. 
\end{assumption}

Note that Assumption~\ref{as:discretization} is easy to satisfy with a minor loss of generality.
While formally relaxing Assumption~\ref{as:same-t*} remains challenging, we numerically demonstrate in Section~\ref{sec:case4} that the strong Pareto improvement is still valid under heterogeneous desired arrival times $t^*_\tau$. 
We also note that the results derived below are based on the simple payment and redistribution rules presented in Section~\ref{sec:karma-transition}. Although we expect they hold for more general karma transition kernels, proving this is beyond the scope of this paper.


\subsection{Fairness definitions}
In the literature, a policy is commonly refereed as ``fair'' if all travelers share the same monetary cost. Under this criterion, in the classic bottleneck model, the equilibrium with a tolled fast lane is considered ``fair'' because commuters have the same total monetary travel cost, i.e., the toll price paid by commuters using the fast lane equals the monetary value of waiting times suffered by commuters on the slow lane.
This notion of ``fairness'' relies on money being a fair, universal and objective measure of need.
However, it is well known that individuals' sensitivity to money varies widely due to wealth, income and other factors, see~\cite{arnott1988schedule,verhoef:2011:synamic_bottleneck,liu2011morning}.
More generally, we believe that defining  fairness with respect to a universal absolute measure is flawed since such a measure does not exist in practice.
Therefore,
we leverage the dynamic \gls{VOT} process to define fairness with respect to a normalized scale based on the individual long-term average \gls{VOT} $\ubar_\tau$.
Intuitively, this new definition implies that a policy is fair if all commuters get an equal opportunity to enter the fast lane on the long term, and that they can do so when they have high \gls{VOT} with respect to their own individual \gls{VOT} scale.


To formalize this notion, at a $\karma$ \gls{SNE}, the \emph{long-term average travel cost} and \emph{normalized long-term average travel cost} for a  type $\tau\in\Gamma$ are defined as follows, 
\begin{subequations}
\begin{align}\label{eq:average-cost}
    \cbar_\tau(d^*,\pi^*) &= -\frac{1}{g_\tau} \: \sum_{u \in \U_\tau} \sum_{k \in \Nat} d^*_\tau[u,k] \: R_\tau[u,k](d^*,\pi^*), \\
    \cbarnorm_\tau(d^*,\pi^*) &= \frac{\cbar_\tau(d^*,\pi^*)}{\ubar_\tau}. \label{eq:norm-average-cost}
\end{align}
\end{subequations}

The long-term average travel cost~\eqref{eq:average-cost} is the negated expected immediate reward $R_\tau$, given in~\eqref{eq:exp-immediate-reward}, at the stationary state distribution of type $\tau$, given by $d^*_\tau / g_\tau$. 
In~\eqref{eq:norm-average-cost} this cost is normalized by the average \gls{VOT} $\ubar_\tau$~\eqref{eq:norm-average-cost}. 

We are now ready to introduce the following two notions of fairness which are defined in terms of the normalized long-term average travel cost.
\begin{definition}[Equitable \gls{SNE}]
\label{def:equity}
A $\karma$ \gls{SNE} $(d^*, \pi^*)$ is said to be equitable if it satisfies
\begin{align}
\cbarnorm_\tau(d^*, \pi^*) = \cbarnorm_{\tau'}(d^*, \pi^*),\;\text{for all } \tau,\tau'\in\Gamma. \label{eq:eq-SNE-pi}
\end{align}
\end{definition}

\begin{definition}[Strong Pareto improvement]
    Let $\{\cbar^\textup{\;n,\texttt{BM}}_\tau\}_{\tau\in\Gamma}$ be the set of normalized long-term average travel costs evaluated at a benchmark equilibrium. 
    A $\karma$ \gls{SNE} $(d^*,\pi^*)$ is said to satisfy the strong Pareto improvement property with respect to the benchmark if, for all $\tau\in\Gamma$, it satisfies
    \begin{align}\label{eq:strong-pareto}
        \cbarnorm_\tau(d^*,\pi^*) < \cbar^\textup{\;n,\texttt{BM}}_\tau.
    \end{align}
\end{definition}

\subsection{Equity with respect to heterogeneous income}
\label{sec:equity}

Let us first define the heterogeneous VOT processes that indicate different income levels.
\begin{definition}[Income-heterogeneous VOT process] 
    Let  $\U_0 = \{u_1,\dots,u_U\}$ be the set of different urgency levels with transition matrix $\phi_0$. Then, the VOT processes are said to be income-heterogeneous if for each type $\tau\in\Gamma$, there is some $\lambda_\tau>0$ such that
    \begin{align}
        &\U_\tau = \{\lambda_\tau u_1,\dots,\lambda_\tau u_U\},\\
        &\phi_\tau[u^+ = \lambda_\tau u_j| u = \lambda_\tau u_i] = \phi_0[u^+ =  u_j| u = u_i],\;\textup{for all } u_i,u_j \in\U_0. 
    \end{align}
\end{definition}

In words, this means that all commuter types essentially have the same \gls{VOT} process up to a difference in the scaling factors $\lambda_\tau$, which represent the heterogeneous income levels of the commuters.
For example, for two income types $\tau = \textup{``}\low\textup{''}$ and $\tau = \textup{``}\high\textup{''}$ with $\lambda_\high > \lambda_\low$, we have $\ubar_\high > \ubar_\low$ corresponding to the \gls{VOT} heterogeneity used in classical monetary bottleneck models~\cite{arnott1988schedule,verhoef:2011:synamic_bottleneck,liu2011morning}.

\begin{proposition}[Equity with respect to heterogeneous income]
\label{prop:equity}
    Under Assumption~\ref{as:same-t*}, an equitable \gls{SNE} exists in a $\karma$ system with income-heterogeneous \gls{VOT} processes.
\end{proposition}

\begin{proof}
    The proof is included in Appendix~\ref{proof:equity}.
\end{proof}

Proposition~\ref{prop:equity} guarantees the existence of an equitable \gls{SNE} rather than asserting the stronger notion that all \gls{SNE} are equitable.
The latter would hold if the \gls{SNE} is unique, which remains an open problem in general \gls{DPG}s.
However, for any two types $\tau = \low$ and $\tau = \high$, even if $\pi^*_\low$ and $\pi^*_\high$ are different at a \gls{SNE} $(d^*,\pi^*)$, by virute of their optimality they must achieve the same
normalized discounted infinite-horizon reward when commuters of both types start from the same state (refer to~\eqref{eq:value-function} for the definition of the discounted infinite horizon reward).
In general, this does not imply an equitable \gls{SNE} because the different policies could lead to different stationary state distributions ($d^*_\low$ and $d^*_\high$) and in turn, different $\cbarnorm_\tau(d^*, \pi^*)$. 
However, with a sufficiently low discounting (i.e., $\delta\to1$), the difference vanishes. In other words, any possible inequity is attributed to the short-sightedness of commuters. This result is formally stated in the following proposition. 


\begin{proposition}
\label{prop:equity-no-discount}
    Under Assumption~\ref{as:same-t*}, any $\karma$ \gls{SNE} with income-heterogeneous VOT processes satisfies
    \begin{equation}
        \lim_{\delta \rightarrow 1} \left[\cbarnorm_\tau(d^*,\pi^*) - \cbarnorm_{\tau'}(d^*,\pi^*)\right] = 0, \;\textup{for all } \tau,\tau'\in\Gamma.
    \end{equation}
\end{proposition}

\begin{proof}
    The proof is included in Appendix~\ref{proof:equity-no-discount}.
\end{proof}

Interestingly, Proposition~\ref{prop:equity-no-discount} implies that equity may be improved by making commuters more aware of their possible future reward, e.g., via daily feedback on their behaviour.

\subsection{Strong Pareto improvement with respect to no policy intervention}
\label{sec:pareto-improvement}

We next consider general heterogeneity in the \gls{VOT} process.
In this case, one would not expect the \gls{SNE} to be equitable since commuters face different decision-making problems, which lead to different strategies on how to spend karma. 
Nonetheless, we  prove that all commuters strictly benefit from $\karma$ compared to the nominal case without policy intervention, denoted by $\nom$ hereafter.

\begin{proposition}
    \label{prop:strict-pareto}
    Under Assumptions~\ref{as:same-t*} and \ref{as:discretization}, all $\karma$ SNE $(d^*,\pi^*)$ satisfy strong Pareto improvement with respect to no policy intervention ($\nom$), i.e., 
    \begin{align}
        \cbarnorm_\tau(d^*,\pi^*) < c^*,\;\textup{for all } \tau\in\Gamma,
    \end{align}
    where $c^*$ is the normalized equilibrium cost under $\nom$, given in \eqref{eq:c-star}. 
\end{proposition}

The formal proof of Proposition~\ref{prop:strict-pareto} is included in Appendix~\ref{proof:strict-pareto} and relies on a series of preparatory results which are included in Appendix~\ref{sec:lemmas}.
We provide here the main intuition {behind} the proof.
Due to the structure of the bottleneck problem it can be shown that at any $\karma$ \gls{SNE}, the normalized immediate cost experienced by any commuter is at most $c^*$ (Lemma~\ref{lem:weak-Pareto}), while it is strictly less than $c^*$ for commuters using the fast lane. Moreover, it can be shown that $\karma$ ensures that all commuters get a chance to use the fast lane in the long run, which is also the optimal choice for them.
It follows that a strong Pareto improvement is guaranteed at a $\karma$ \gls{SNE}.

\section{Numerical experiments}
\label{sec:numerical-analysis}
In this section, we compare $\karma$ with the nominal bottleneck equilibrium ($\nom$) and the optimal tolling on the fast lane (denoted by $\toll$) in several case studies that are constructed with different user heterogeneity. The results further highlight the novel features of $\karma$ presented above.
Specifically, in Section~\ref{sec:measures}, we introduce the performance measures as well as the default model parameters. Section~\ref{sec:case1} investigates the performance of $\karma$ with homogeneous commuters with the objective to demonstrate that $\karma$ achieves the same performance as $\toll$ in reducing the congestion and the average user travel cost. Then, in Section~\ref{sec:case2}, we consider two types of commuters with different constant \gls{VOT}. This mimics the classical setting of the bottleneck model with heterogeneous income levels. The objective here is to showcase that $\karma$ is equitable with respect to income (Proposition~\ref{prop:equity}) and contrast it to $\toll$ that exhibits severe inequity.
In Section~\ref{sec:case3}, we validate the strong Pareto improvement property of $\karma$ under the general \gls{VOT} process heterogeneity and show that $\toll$ does not satisfy this property even if all commuter types have the same average \gls{VOT}.
Finally, in Section~\ref{sec:case4}, we introduce heterogeneity in the desired arrival times to explore the potential of $\karma$ beyond our theoretical analysis and show how also in this case it manages to 
reduce congestion in a fair manner.
Note that in all experiments conducted in this section, karma is redistributed uniformly among all commuters, while the more sophisticated redistribution schemes are investigated in the next section. 


\subsection{Performance measures and benchmark}
\label{sec:measures}
The performance measures can be divided into two categories: system level and user-type level. Specifically, we investigate the long-term average queuing delay (in unit of minute) and {the normalized long-term average travel cost (see \eqref{eq:norm-average-cost})}
at the equilibrium state. 

\begin{table}[ht]
\caption{Performance measures in three models with homogeneous desired arrival time.}
\label{tab:measure}
\centering
\renewcommand{\arraystretch}{1.2}
\begin{tabular}{|p{0.17\textwidth}p{0.02\textwidth}|p{0.14\textwidth}|p{0.2\textwidth}|p{0.34\textwidth}|}
\hline
\textbf{Name}                                  &        & \textbf{Benchmark}            & \textbf{Optimal tolling}      & \textbf{Karma mechanism}$^\star$\\
& & ($\nom$) &  ($\toll$) & ($\karma$) \\
\hline
System average queuing delay     & $\bar{t}^q$      & $\frac{c^*}{2\alpha}$           &      $\frac{s_\textup{slow}}{s} \: \frac{c^*}{2\alpha}$     &     $\sum_{t,b} \nu[t,b] \sum_o \psi^\epsilon[o|t,b] \: t^q[t,o]$                       \\
\hline
System average travel cost       & $\cbarnorm$        & $c^*$ &                  $\sum_\tau g_\tau \: \cbarnorm_\tau$     &    $\sum_\tau g_\tau \: \cbarnorm_\tau$   \\
\hline
Type average queuing delay & $\bar{t}^q_\tau$ &    $\frac{c^*}{2\alpha}$              &    $\sum_u \Prob_\tau[u] \: \bar{t}^q[u]$    & 
$\sum_{t,b}\nu_\tau[t,b]  \sum_o \psi^\epsilon[o|t,b] \: t^q[t,o]$ \\
\hline
Type average travel cost & $\cbarnorm_\tau$   &         $c^*$              &    $\frac{1}{\ubar_\tau} \sum_u \Prob_\tau[u] \: c^*[u]$   &    
 $-\frac{1}{g_\tau \, \ubar_\tau} \sum_{u,k} d^*_\tau[u,k] \: R_\tau[u,k]$ \\
\hline
\multicolumn{5}{|l|}{$^\star$All measures are computed at the \gls{SNE}.}\\
\hline
\end{tabular}
\end{table}

Table~\ref{tab:measure} summarizes the performance measures with homogeneous desired arrival time, i.e., $t^*_\tau\equiv t^*$, for all $ \tau\in\Gamma$. 
Most of the expressions are derived from the literature \cite{arnott1990departure,xiao2013managing}. Specifically, $c^*=\frac{\beta\gamma}{\beta+\gamma}\frac{N}{s}$ is the normalized equilibrium cost under $\nom$, and $c^*[u]$ ($\bar{t}^q[u]$) denotes the travel cost (queuing delay) for commuters with \gls{VOT} $u$ under $\toll$. See Appendix~\ref{sec:bottleneck} for their specifications. 
The measures under $\karma$, on the other hand, are all computed using the variables defined in Section~\ref{sec:karma-bottleneck}, except for $\nu_\tau[t,b](d^*,\pi^*) := \frac{1}{g_\tau} \sum_{u \in \U_\tau} \sum_{k \in \Nat} d^*_\tau[u,k] \: \pi^*_\tau[t,b \mid u,k]$ that, similarly to \eqref{eq:prob-outcome}, specifies type $\tau$ commuters' probability of taking action $[t,b]$.



The default values of model parameters are reported in Table~\ref{tab:default_value}. Besides, the starting time of the congestion period in all case studies is shifted to time zero (i.e., $\tstart=0$), and in Section~\ref{sec:case1} through Section~\ref{sec:case3}, the homogeneous arrival time is set to $t^*=120$ min.

\begin{table}[ht]
\caption{Default values of model parameters.}
\label{tab:default_value}
\centering
\renewcommand{\arraystretch}{1.2}
\begin{tabular}{|p{0.4\textwidth}|p{0.15\textwidth}|p{0.15\textwidth}|p{0.15\textwidth}|}
\hline
\textbf{Name}                           & \textbf{Notation}         & \textbf{Unit}    & \textbf{Value}                      \\ \hline
Number of commuters            &  $N$                &     veh    & 9000                       \\ \hline
Bottleneck capacity            & $s$        & veh/min    & 60 \\
- fast lane                    & $s_\textup{fast}$      &    & 12                         \\ 
- low lane                     & $s_\textup{slow}$      &    & 48                         \\ \hline
Length of discrete time step   & $\Delta$ &  min       & 15                         \\ \hline
Normalized delay penalties                 &                  & 1/hour &                            \\ 
- queuing delay                &  $\alpha$        &  & 6.4                        \\ 
- early arrival                &  $\beta$        &  & 4                          \\ 
- late arrival                 &  $\gamma$         &  & 16                         \\ \hline
Discount factor in MDP               &  $\delta$          &         &  0.99                          \\ \hline
Approximation parameter &  $\epsilon$         &         & $10^{-4}$ 
\\ \hline
Average karma per commuter     &  $\bar{k}$       &         & 10                         \\
\hline
\end{tabular}
\end{table}

\subsection{{$\karma$ is as efficient as optimal tolling}}
\label{sec:case1}
The purpose of this case study is to a) provide insights into the strategic behaviors of commuters under $\karma$, 
and b) demonstrate that $\karma$ closely approaches the efficiency of $\toll$ in a decentralized manner, i.e., without exchange of private information like \gls{VOT} and desired arrival time.
To this end, we consider homogeneous commuters with two possible \gls{VOT} levels. Specifically, commuters have low \gls{VOT} ($\ulow=1$) for $80\%$ of the time and high \gls{VOT} ($\uhigh=6$) for $20\%$ of the time. 
The corresponding \gls{VOT} Markov chain is given by $\phi = \begin{pmatrix} 0.8 & 0.2 \\ 0.8 & 0.2 \end{pmatrix}$.

Figure~\ref{fig:results-case-1}a illustrates the equilibrium policy $\pi^*$ under $\karma$. 
Specifically, each subplot corresponds to one combination of departure time and \gls{VOT} level. The bidding probability given the level of karma (on the horizontal axis) is depicted as the intensity of red color (the higher probability the darker the  color), while the infeasible bids are displayed in gray. The threshold bid $b^*$ at each departure time is also plotted for reference.
As desired for an efficient allocation of the fast lane, low-\gls{VOT} commuters tend to bid less than $b^*$ in order to save karma for the future, whereas high-\gls{VOT} commuters bid $b^*$ in order to enter the fast lane without overpaying for it.
With more karma in hand, the high-\gls{VOT} commuters tend to depart closer to $t^*$ and bid more for the fast lane.
Figure~\ref{fig:results-case-1}b illustrates the associated stationary karma distribution derived from $d^*$.

Figure~\ref{fig:results-case-1}c plots $b^*$ over time, along with the optimal tolling price under $\toll$. Despite that karma and money are incomparable, the trend of $b^*$ over time resembles the optimal toll $p^*$ remarkably well: $b^*$ is the highest at the most competitive departure time $t^*$ and decreases linearly away from it. It is worth noting that $b^*$ arises in a decentralized manner whereas $p^*$ is designed by a central controller.

The similarity between $\karma$ and $\toll$ is also demonstrated in their aggregate outcomes. As shown in Figure~\ref{fig:results-case-1}d, they lead to almost the same departure and congestion patterns. Specifically, the departure rate on the slow lane is constant and higher than its capacity before the peak time $t_\textup{peak}=45$ min and then drops to another constant value below the capacity till the end of the rush hours. Moreover, consistent with Figure~\ref{fig:results-case-1}a the fast lane is largely occupied by high-\gls{VOT} commuters under $\karma$, also in line with the result under $\toll$. In other words, $\karma$ enables the most urgent commuters to use the fast lane and thus allocates it efficiently.

Compared to the non-controlled benchmark $\nom$, both $\karma$ and $\toll$ help reduce the average queuing delay and travel cost\footnote{In $\karma$, a linear interpolation of the queuing delays was performed in order to counter-act the effect of the course of  time discretization, see e.g., \cite{ramadurai2010linear}.}. As shown in Figure~\ref{fig:results-case-1}e, their improvements are also quite close to each other ($20\%$ reduction in queuing delay and $30\%$ reduction in travel cost). Recall that $\toll$ leads to the system optimum when only the fast lane can be regulated. Therefore we conclude that $\karma$ closely achieves the system optimum with homogeneous commuters.

\begin{minipage}[t]{0.35 \textwidth}	
    \vspace{0pt}
    \centering
    \footnotesize
    
    \def \t {1}
    \def \tleft {30.23203pt}
    \def \tright {10.40065pt}
    \def \tleftt {15.36607pt}
    \def \trightt {11.47704pt}
    \includegraphics[trim={\tleft{} 0 \tright{} 0}]{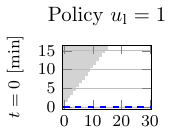}
    \hfil
    \includegraphics[trim={\tleftt{} 0 \trightt{} 0}]{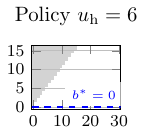}
    
    \def \t {2}
    \def \tleft {29.72046pt}
    \def \tright {7.39874pt}
    \def \tleftt {16.53252pt}
    \def \trightt {7.39874pt}
    \includegraphics[trim={\tleft{} 0 \tright{} 0}]{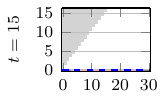}
    \hfil
    \includegraphics[trim={\tleftt{} 0 \trightt{} 0}]{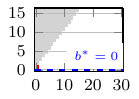}
    
    \def \t {3}
    \includegraphics[trim={\tleft{} 0 \tright{} 0}]{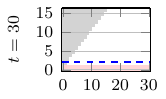}
    \hfil
    \includegraphics[trim={\tleftt{} 0 \trightt{} 0}]{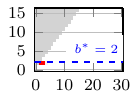}
    
    \def \t {4}
    \includegraphics[trim={\tleft{} 0 \tright{} 0}]{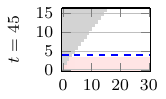}
    \hfil
    \includegraphics[trim={\tleftt{} 0 \trightt{} 0}]{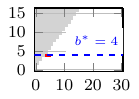}
    
    \def \t {5}
    \includegraphics[trim={\tleft{} 0 \tright{} 0}]{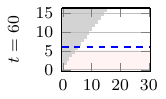}
    \hfil
    \includegraphics[trim={\tleftt{} 0 \trightt{} 0}]{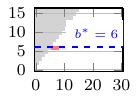}
    
    \def \t {6}
    \includegraphics[trim={\tleft{} 0 \tright{} 0}]{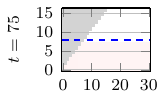}
    \hfil
    \includegraphics[trim={\tleftt{} 0 \trightt{} 0}]{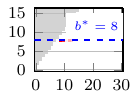}
    
    \def \t {7}
    \includegraphics[trim={\tleft{} 0 \tright{} 0}]{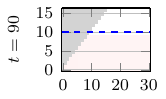}
    \hfil
    \includegraphics[trim={\tleftt{} 0 \trightt{} 0}]{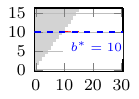}
    
    \def \t {8}
    \includegraphics[trim={\tleft{} 0 \tright{} 0}]{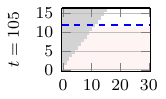}
    \hfil
    \includegraphics[trim={\tleftt{} 0 \trightt{} 0}]{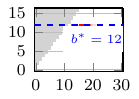}
    
    \def \t {9}
    \def \tleft {30.22159pt}
    \includegraphics[trim={\tleft{} 0 \tright{} 0}]{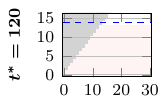}
    \hfil
    \includegraphics[trim={\tleftt{} 0 \trightt{} 0}]{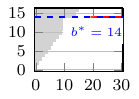}
    
    \def \t {10}
    \def \tleft {29.72046pt}
    \includegraphics[trim={\tleft{} 0 \tright{} 0}]{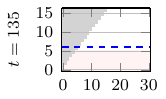}
    \hfil
    \includegraphics[trim={\tleftt{} 0 \trightt{} 0}]{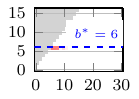}
    
    \def \t {11}
    \includegraphics[trim={\tleft{} 0 \tright{} 0}]{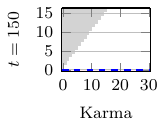}
    \hfil
    \includegraphics[trim={\tleftt{} 0 \trightt{} 0}]{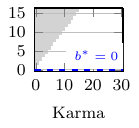}
    
    (a)
\end{minipage}
\hspace{-5mm}
\vline
\hspace{1mm}
\begin{minipage}[t]{0.6 \textwidth}
    \vspace{0pt}
    \centering
    \footnotesize

    \def \tleft {31.50934pt}
    \def \tright {17.99263pt}
    \def \tleftt {34.56491pt}
    \def \trightt {41.17612pt}
    \includegraphics[trim={\tleft{} 0 \tright{} 0}]{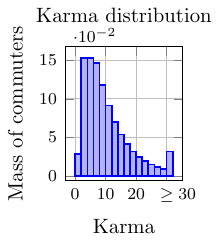}
    \hfil
    \includegraphics[trim={\tleftt{} 0 \trightt{} 0}]{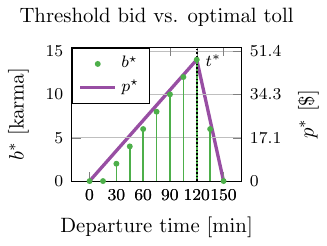}
    
    (b) \hspace{36mm} (c)
    \medskip
    \hrule
    \smallskip
    
    \def \tleft {35.7594pt}
    \def \tright {23.29237pt}
    \def \tleftt {16.53252pt}
    \def \trightt {7.20901pt}
    \includegraphics[trim={\tleft{} 0 \tright{} 0}]{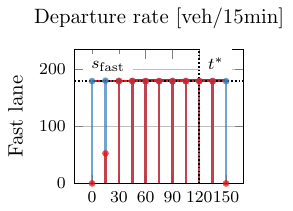}
    \hfil
    \includegraphics[trim={\tleftt{} 0 \trightt{} 0}]{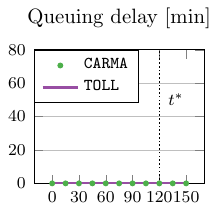}

    \def \tleft {33.42651pt}
    \def \tright {8.80634pt}
    \def \trightt {9.38956pt}
    \includegraphics[trim={\tleft{} 0 \tright{} 0}]{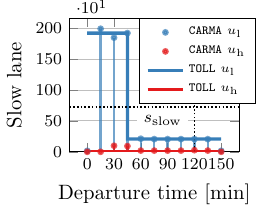}
    \hfil
    \includegraphics[trim={\tleftt{} 0 \trightt{} 0}]{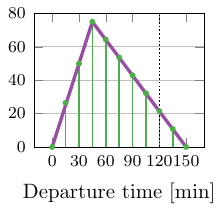}

    (d)
    \medskip
    \hrule
    \smallskip
    
    \def \tleft {34.58792pt}
    \def \tright {12.5009pt}
    \def \tleftt {28.87482pt}
    \def \trightt {4.45932pt}
    \includegraphics[trim={\tleft{} 0 \tright{} 0}]{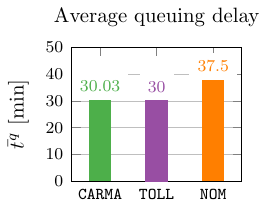}
    \hfil
    \includegraphics[trim={\tleftt{} 0 \trightt{} 0}]{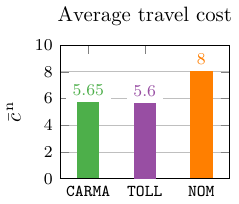}
    
    (e)
    \captionof{figure}{Results for homogeneous commuters.
    (a) \acrfull{SNE} policy.
    (b) \acrshort{SNE} karma distribution.
    (c) Threshold bid $b^\star$ in $\karma$ versus optimal toll price $p^*$ in $\toll$.
    (d) Left: departure rate in $\karma$ versus $\toll$.
    Right: queuing delay over time.
    (e) Performance comparison of $\karma$ versus $\toll$ and $\nom$.
    }
    \label{fig:results-case-1}
\end{minipage}

\subsection{{$\karma$ is equitable with respect to income but optimal tolling is not}}
\label{sec:case2}

{The purpose of this case study is to validate Proposition~\ref{prop:equity} and compare the sensitivity to income of $\karma$ and $\toll$.}
To this end, we model two income types with different constant \gls{VOT}: 
the \emph{low-income} type $\tau=\low$ with $\U_\low=\{1\}$, and the \emph{high-income} type $\tau=\high$ with $\U_\high=\{6\}$.
The low-income commuters occupy $80\%$ of the population and the high-income commuters the remaining $20\%$.


As can be seen from Figure~\ref{fig:results-case-2}a, $\karma$ and $\toll$ achieve the same aggregate departure and congestion patterns. The total departure rates as well as the queuing delays are identical to the homogeneous case; as per Figure~\ref{fig:results-case-1}d.
However, the two models behave quite differently when we compare the results by commuter type. 
Under $\toll$, the fast lane is fully occupied by high-income commuters whereas, under $\karma$, the fast lane capacity is split between the two types in the same ratio as their proportions in the population.
In the former case, the optimal toll price essentially discriminates against the low-income commuters from using the fast lane.
As a result, $\toll$ leads to a serious equity issue in terms of both queuing delay and travel cost
as demonstrated in Figure~\ref{fig:results-case-2}b. Specifically, the high-income commuters enjoy no queuing delay at all while the low-income commuters get stuck in the queue for 37 min on average. Their average travel cost also doubles compared to the high-income commuters. 

In contrast, under $\karma$, the two types of commuters share the same queuing delay and travel cost, as predicted by Proposition~\ref{prop:equity}. Figure~\ref{fig:results-case-2}c provides further insight. At the \gls{SNE}, both types of commuters have the same distribution of karma. Accordingly, they have equal opportunity to bid for the fast lane.
In fact, their departure and bidding strategies are exactly the same (see the top two subplots in Figure~\ref{fig:results-case-2}c for an example at $t=t^*$) because they essentially face the same optimization problem (up to a constant difference in the scaling of costs).

\begin{minipage}[t]{0.45\textwidth}
\vspace{0pt}
\centering
\footnotesize

\def \tleft {33.42651pt}
\def \tright {21.73213pt}
\def \tleftt {15.36607pt}
\def \trightt {5.64877pt}
\includegraphics[trim={\tleft{} 0 \tright{} 0}]{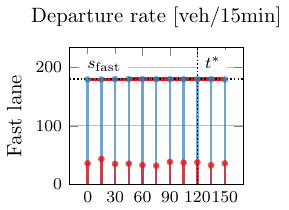}
\hfill
\includegraphics[trim={\tleftt{} 0 \trightt{} 0}]{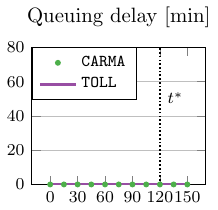}
\hspace{1mm}

\def \tleft {33.42651pt}
\def \tright {12.55478pt}
\def \tleftt {15.36607pt}
\def \trightt {7.82932pt}
\includegraphics[trim={\tleft{} 0 \tright{} 0}]{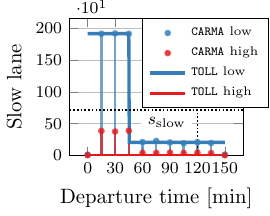}
\hfill
\includegraphics[trim={\tleftt{} 0 \trightt{} 0}]{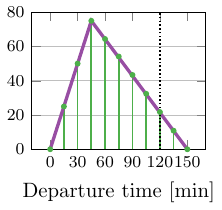}
\hspace{1mm}

(a)
\medskip
\hrule
\smallskip

\def \tleft {32.25504pt}
\def \tright {10.94264pt}
\def \tleftt {31.34698pt}
\def \trightt {3.48428pt}
\includegraphics[trim={\tleft{} 0 \tright{} 0}]{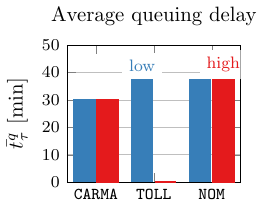}
\hfill
\includegraphics[trim={\tleftt{} 0 \trightt{} 0}]{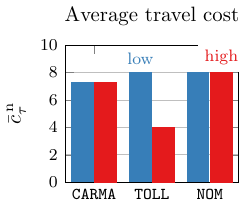}
\hspace{1mm}

\centering
(b)
\end{minipage}
\vline
\hspace{1mm}
\begin{minipage}[t]{0.45\textwidth}
\vspace{0pt}
\raggedleft
\footnotesize

\def \t {9}

\def \tleft {31.12091pt}
\def \tright {6.0786pt}
\def \tleftt {6.86595pt}
\def \trightt {6.0786pt}
\includegraphics[trim={\tleft{} 0 \tright{} 0}]{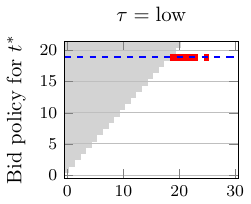}
\hspace{5mm}
\includegraphics[trim={\tleftt{} 0 \trightt{} 0}]{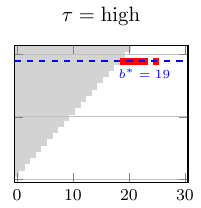}

\def \tleft {29.17645pt}
\def \tright {6.1649pt}
\def \tleftt {6.86595pt}
\def \trightt {6.1649pt}
\includegraphics[trim={\tleft{} 0 \tright{} 0}]{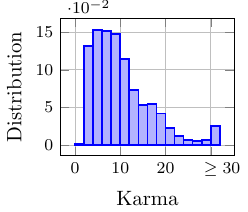}
\hspace{5mm}
\includegraphics[trim={\tleftt{} 0 \trightt{} 0}]{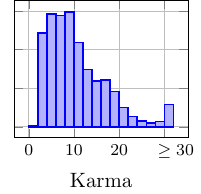}

\centering
(c)
\captionof{figure}{Results for heterogeneous income levels.
(a) Left: departure rate in $\karma$ versus $\toll$.
Right: queuing delay over time.
(b) Fairness of $\karma$ versus $\toll$ and $\nom$.
(c) \gls{SNE} with equilibrium policy at $t^*=120~\text{min}$ (top) and stationary karma distribution (bottom) per type $\tau$.}
\label{fig:results-case-2}
\end{minipage}

\bigskip

Another important observation from Figure~\ref{fig:results-case-2}b is that only high-income commuters benefit from $\toll$, while the low-income commuters remain the same as in $\nom$. Differently, in line with Proposition~\ref{prop:strict-pareto}, all commuters enjoy a shorter queuing delay and a lower travel cost thanks to $\karma$. Furthermore, we note that the fairness of $\karma$ does not come at the cost of efficiency, as we still have the same reduction in congestion with respect to $\nom$, i.e., same average queuing delay as in Figure~\ref{fig:results-case-1}e.

\subsection{{$\karma$ satisfies strong Pareto improvement but optimal tolling does not}} 
\label{sec:case3}

{The purpose of this case study is to validate Proposition~\ref{prop:strict-pareto} and compare $\karma$ to $\toll$ under general \gls{VOT} process heterogeneity.}
Consider four commuter types each constituting a quarter of the population. We assume 
all types share the same average \gls{VOT} $\ubar_\tau=2$, for all $\tau\in\Gamma=\{1,2,3,4\}$, but differ in the magnitude and frequency of the high \gls{VOT} level. 
The type-specific \gls{VOT} processes are summarized in Table~\ref{tab:heter-urgency}.

\begin{table}[ht]
\caption{\gls{VOT} processes of commuters by type.}
\label{tab:heter-urgency}
\centering
\footnotesize
\renewcommand{\arraystretch}{1.2}
\begin{tabular}{|p{0.15\textwidth}|p{0.15\textwidth}|p{0.15\textwidth}|p{0.2\textwidth}|p{0.15\textwidth}|}
\hline
\textbf{\gls{VOT} \mbox{process}} & \textbf{Low \gls{VOT}} ($\ulow$) & \textbf{High \gls{VOT}} ($\uhigh$) & \textbf{Probability of high \gls{VOT}}& \textbf{\gls{VOT} spread} \\
\hline
$\phi_1$ & 1 & 11 & $10\%$ & 10\\
\hline
$\phi_2$ & 1 & 6 & $20\%$& 5\\
\hline
$\phi_3$ & 1 & 3 & $50\%$& 2\\
\hline
$\phi_4$ & 2 & 2 & $100\%$& 0 \\
\hline
\end{tabular}
\end{table}

Note that from Type $1$ to Type $4$, the commuters have a decreasing spread in their \gls{VOT} levels.
Type $1$ has the widest spread, incurring the very high \gls{VOT} $\uhigh=11$ on relatively rare occasions.
On the other hand, Type $4$ has the constant \gls{VOT} $u=2$ all the time.


Similar to the previous two case studies, we first compare the departure patterns in $\karma$ and $\toll$. The left column in Figure~\ref{fig:results-case-3}a illustrates the departure rates under $\karma$, where the commuter type and \gls{VOT} level are plotted in different colors (each hue represents one type and the high \gls{VOT} levels are displayed with higher saturation). 
The departure rates under $\toll$ are plotted in the right column and they are solely determined by realized \gls{VOT} levels on a single day. 
In line with classical results~\cite{liu2011morning}, there is a perfect separation in the departure time intervals: commuters with higher \gls{VOT} levels ($u=3,6,11$) enter the fast lane and depart closer to $t^*$, while those with lower \gls{VOT} levels ($u=1,2$) only use the slow lane. 

As expected, the aggregate departure pattern under $\karma$ is exactly the same as that under $\toll$. However, they are drastically different when analyzed by commuter type.  
Under $\karma$, commuters on the fast lane are largely heterogeneous, and the departure intervals of different types and \gls{VOT} levels are partially overlapping. Nevertheless, we can observe that the departures on the fast lane at $t^*$ are dominated by commuters with the highest \gls{VOT} (mostly $u=11$),  while those with the lowest urgency ($u=1$) tend to use the slow lane. Even if they enter the fast lane, they would choose a departure time far from $t^*$ to save their karma. 

Figure~\ref{fig:results-case-3}b plots the average queuing delay and travel cost by commuter type in the three schemes.
This gives a concrete example where $\toll$ fails to satisfy the strong Pareto improvement property despite all commuters having the same average \gls{VOT}:
commuters in Type $4$ (with constant \gls{VOT} $u=2$) do not benefit from the tolling scheme at all because they never use the fast lane. On the other hand, commuters in Type $3$ have the largest reduction in queuing delay while those in Type $1$ enjoy the most travel cost saving.

In contrast to $\toll$ and in line with Proposition~\ref{prop:strict-pareto}, $\karma$ strictly benefits all commuters.
Moreover, it achieves a considerable equalizing effect in the sense of similar queuing delay and a narrow range of travel cost.
However, unlike the case of heterogeneous incomes (Figure~\ref{fig:results-case-2}b), strict equality among types is not obtained in this case. This is expected because, despite of sharing the same average \gls{VOT}, the heterogeneous \gls{VOT} processes, particularly the different \gls{VOT} spreads,  lead to different value functions in each commuter's \gls{MDP}
and, in turn, different optimal policies. 

Another interesting observation from Figure~\ref{fig:results-case-3}b is that $\karma$ encourages \emph{yielding}: those suffering from a longer queue actually enjoy a lower travel cost. In other words, it is not optimal to aggressively bid for the fast lane.
Also note that the travel cost under $\karma$ decreases with the frequency of high-urgency events. This result implies that, if commuters are free to choose their \gls{VOT} processes to minimize their average travel cost, they would tend to be cautious and yielding most of the time and be urgent only sparingly. Therefore, $\karma$ fosters a more collaborative environment where commuters carefully assess whether they truly need access to the fast lane. 

To better illustrate the commuters' bidding policies under $\karma$, we plot the karma distribution at the \gls{SNE} in Figure~\ref{fig:results-case-3}c. It can be found that the most yielding commuters (Type 1) hold more karma on average than others, and the average karma per commuter decreases with the \gls{VOT} spread. This result is also expected because commuters who anticipate high-urgency events in the future are more likely to yield on low-urgency days to reserve karma. On the other hand, if the variation in the daily \gls{VOT} level is small, commuters tend to avoid congestion whenever they have enough karma.

\begin{minipage}[t]{0.45\textwidth}
\vspace{0pt}
\centering
\footnotesize

\def \tleft {33.42651pt}
\def \tright {0.89734pt}
\def \tleftt {19.61613pt}
\def \trightt {0.89734pt}
\includegraphics[trim={\tleft{} 0 \tright{} 0}]{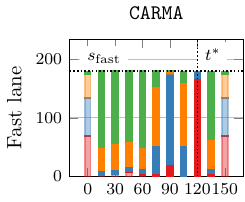}
\hfill
\includegraphics[trim={\tleftt{} 0 \trightt{} 0}]{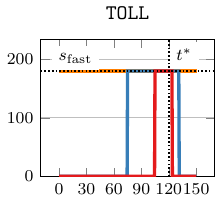}
\hspace{3mm}

\def \tleft {33.42651pt}
\def \tright {8.77625pt}
\def \tbottom {81.8126pt}
\def \tleftt {19.61613pt}
\def \trightt {7.82932pt}
\def \tbottomm {66.06789pt}
\includegraphics[trim={\tleft{} \tbottom{} \tright{} 0}]{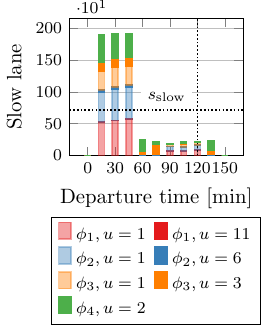}
\hfill
\includegraphics[trim={\tleftt{} \tbottomm{} \trightt{} 0}]{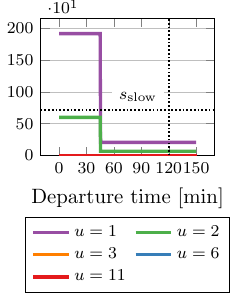}
\hspace{3mm}

\vspace{\tbottom{}}

(a)
\medskip
\hrule
\smallskip
\centering

\def \tleft {32.25504pt}
\def \tright {10.94264pt}
\def \tleftt {31.34698pt}
\def \trightt {3.48428pt}
\includegraphics[trim={\tleft{} 0 \tright{} 0}]{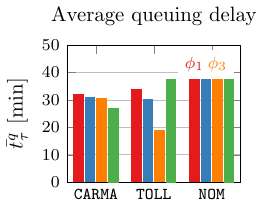}
\hfill
\includegraphics[trim={\tleftt{} 0 \trightt{} 0}]{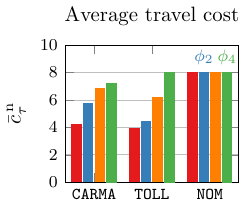}
\hspace{1mm}

(b)

\end{minipage}
\vline
\hspace{1mm}
\begin{minipage}[t]{0.45\textwidth}
\vspace{0pt}
\raggedleft
\footnotesize

\def \tleft {29.17645pt}
\def \tright {6.1649pt}
\def \tleftt {6.86595pt}
\def \trightt {6.1649pt}
\includegraphics[trim={\tleft{} 0 \tright{} 0}]{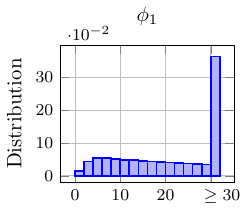}
\hspace{5mm}
\includegraphics[trim={\tleftt{} 0 \trightt{} 0}]{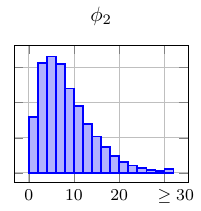}

\includegraphics[trim={\tleft{} 0 \tright{} 0}]{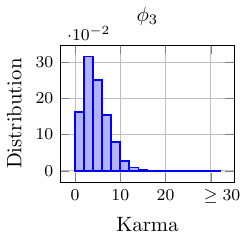}
\hspace{5mm}
\includegraphics[trim={\tleftt{} 0 \trightt{} 0}]{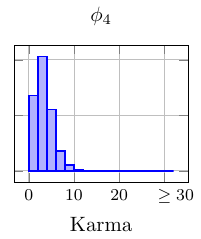}

\centering
(c)
\captionof{figure}{Results for heterogeneous \gls{VOT} processes.
(a) Departure rate in $\karma$ (left) versus $\toll$ (right) (in veh/15min).
(b) Fairness of $\karma$ versus $\toll$ and $\nom$.
(c) \gls{SNE} stationary karma distribution per \gls{VOT} process.}
\label{fig:results-case-3}
\end{minipage}
\bigskip

\subsection{{$\karma$ adapts to heterogeneous desired arrival times}}
\label{sec:case4}

Finally, we investigate the performance of $\karma$ under heterogeneity in the desired arrival times. Specifically, we consider two commuter types $\tau \in \Gamma=\{1,2\}$ with the same VOT process in Section~\ref{sec:case1} but differing in their desired arrival times $t^*_\tau$. 
The first type has $t^*_1=120$ min as before, while the second type has $t^*_2$ according to the following three settings, which yield different aggregate departure patterns: 
\begin{enumerate}
    \item \emph{Single peak:} $t^*_2=60$ min and $g_2=20\%$; \label{case:single-peak}

    \item \emph{Double peak:}  $t^*_2=60$ min and $g_2=50\%$; \label{case:double-peak}

    \item \emph{Pre-congestion:}  $t^*_2=0$ min and $g_2=2\%$. \label{case:uncontested}
\end{enumerate}

Table~\ref{tab:measure-hetero-t*} in Appendix~\ref{sec:bottleneck-hetero-t*} reports the performance measures under $\nom$ and $\toll$ in each study case, while those under $\karma$ remain the same as in Table~\ref{tab:measure}. The detailed derivation is also included there.

\begin{figure}[!tb]
\centering

\def \tleft {33.4538pt}
\def \tright {3.56674pt}
\def \tbottom {13.18794pt}
\def \tleftt {16.53252pt}
\def \trightt {4.44174pt}
\def \tbottomm {13.18794pt}
\def \tlefttt {16.53252pt}
\def \trighttt {10.12231pt}
\def \tbottommm {13.18794pt}
\def \tleftttt {16.53252pt}
\def \trightttt {6.74812pt}
\def \tbottommmm {14.73018pt}
\includegraphics[trim={\tleft{} \tbottom{} \tright{} 0}]{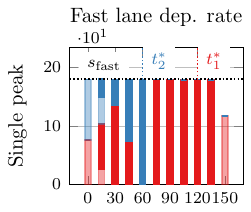}
\hfil
\includegraphics[trim={\tleftt{} \tbottomm{} \trightt{} 0}]{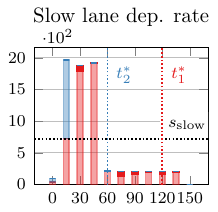}
\hfil
\includegraphics[trim={\tlefttt{} \tbottommm{} \trighttt{} 0}]{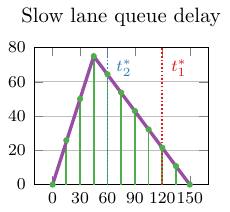}
\hfil
\includegraphics[trim={\tleftttt{} \tbottommmm{} \trightttt{} 0}]{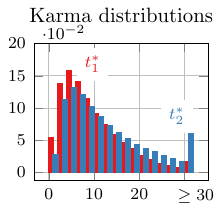}

\vspace{\tbottommmm{}}

\def \tright {3.56674pt}
\def \trightt {1.48087pt}
\def \trighttt {1.48087pt}
\includegraphics[trim={\tleft{} \tbottom{} \tright{} 0}]{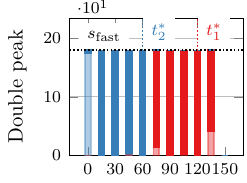}
\hfil
\includegraphics[trim={\tleftt{} \tbottomm{} \trightt{} 0}]{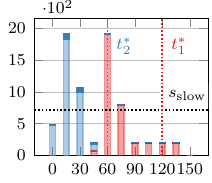}
\hfil
\includegraphics[trim={\tlefttt{} \tbottommm{} \trighttt{} 0}]{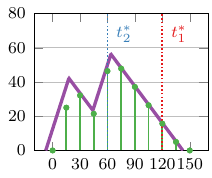}
\hfil
\includegraphics[trim={\tleftttt{} \tbottommmm{} \trightttt{} 0}]{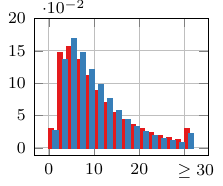}

\vspace{\tbottommmm{}}

\def \tright {1.48087pt}
\def \tbottom {29.99808pt}
\def \tbottomm {29.99808pt}
\def \tbottommm {29.99808pt}
\def \tleftttt {16.53252pt}
\def \trightttt {0.2pt}
\def \tbottommmm {28.05365pt}
\includegraphics[trim={\tleft{} \tbottom{} \tright{} 0}]{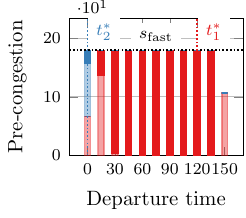}
\hfil
\includegraphics[trim={\tleftt{} \tbottomm{} \trightt{} 0}]{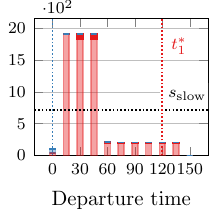}
\hfil
\includegraphics[trim={\tlefttt{} \tbottommm{} \trighttt{} 0}]{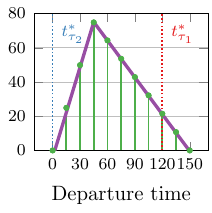}
\hfil
\includegraphics[trim={\tleftttt{} \tbottommmm{} \trightttt{} 0}]{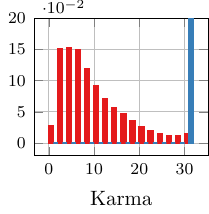}

\vspace{\tbottom{}}

\caption{Results for heterogeneous desired arrival times. From left to right: fast and slow lane departure rates (in veh/15min) per desired arrival time ($t^*_1$ vs. $t^*_2$) and \gls{VOT} (light: $\ulow=1$, dark: $\uhigh=6$); slow lane queuing delays of $\karma$ vs. $\toll$; stationary karma distribution per desired arrival time.}
\label{fig:results-case-4}
\end{figure}

As can be seen in Figure~\ref{fig:results-case-4}, $\karma$ naturally adapts to the heterogeneous desired arrival times meanwhile allocating the fast lane efficiently.
The fast lane is dominated by the high \gls{VOT} commuters, which depart closest to their desired arrival times. In the single peak setting, the Type 2 commuters fully occupy the fast lane at their desired arrival time $t^*_2$ and partially occupy it during the three preceding time steps, which are shared with the Type 1 commuters.
In the double peak setting, the departure time windows on the fast lane are fully separated, with Type 2 departing earlier than Type 1.
Both results are in line with the departure pattern under $\toll$ and thus lead to similar congestion reduction and travel cost saving as reported in Table~\ref{tab:heter-tstar}.

To shed light on $\karma$'s ability to adapt to heterogeneous desired arrival times, the right-most column of Figure~\ref{fig:results-case-4} shows the stationary karma distributions per type.
In the single peak setting, the Type 2 commuters hold slightly more karma on average than the Type 1 commuters, although the difference in the distributions is small.
This is expected because $t^*_2$ is less contested than $t^*_1$.
In the double peak setting, the karma distributions of both types are largely aligned since both desired arrival times are equally contested.
In contrast, the pre-congestion setting presents an interesting corner case.
Since the desired arrival time of the Type 2 commuters is \emph{not contested}, with a zero bid, these commuters are able to arrive at their desired time without any delay. 
Consequently, with uniform karma redistribution, these commuters tend to collect all the karma in the system.
A simple resolution is to set a cap on the karma balance beyond which commuters do not receive any karma in the redistribution. In other words, these commuters are effectively eliminated from the system. Figure~\ref{fig:results-case-4} presents the result with a cap of $\kmax = 30$ (three times the average karma among commuters). 
Comparing the results in Table~\ref{tab:heter-tstar} with those in the homogeneous case (see Figure~\ref{fig:results-case-1}e), we may conclude that the karma cap has a negligible effect on the system performance.

\begin{table}[ht]
\caption{Performance comparison with heterogeneous desired arrival times.}
\label{tab:heter-tstar}
\centering
\begin{tabular}{|l|l|c|c|c|c|c|c|}
\cline{3-8}
\multicolumn{2}{l|}{\multirow{2}{*}{}} & \multicolumn{2}{|l|}{Single peak} & \multicolumn{2}{|l|}{Double peak} & \multicolumn{2}{|l|}{Pre-congestion} \\
\cline{3-8}
\multicolumn{2}{l|}{} & $t^*_1 = 120$ & $t^*_2 = 60$ & $t^*_1 = 120$ & $t^*_2 = 60$ & $t^*_1 = 120$ & $t^*_2 = 0$ \\
\hline
\hline
\multirow{3}{*}{$\bar t^q_\tau$} & \cellcolor{green!25}$\karma$ & \cellcolor{green!25}35.35 & \cellcolor{green!25}8.94 & \cellcolor{green!25}27.96 & \cellcolor{green!25}12.88 & \cellcolor{green!25}30.45 & \cellcolor{green!25}0 \\
\cline{2-8}
 & $\toll$ & 33.75 & 15.00 & 29.0625 & 17.8125 & 29.4 & 0 \\
\cline{2-8}
 & $\nom$ & 42.1875 & 18.75 & 36.3281 & 22.2656 & 36.75 & 0 \\
\hline
\hline
\multirow{3}{*}{$\cbarnorm_\tau$} & \cellcolor{green!25}$\karma$ & \cellcolor{green!25}5.53 & \cellcolor{green!25}2.13 & \cellcolor{green!25}3.62 & \cellcolor{green!25}2.83 & \cellcolor{green!25}5.6 & \cellcolor{green!25}0 \\
\cline{2-8}
& $\toll$ & 5.6 & 2.8 & 4.2 & 3.15 & 5.488 & 0 \\
\cline{2-8}
& $\nom$ & 8 & 4 & 6 & 4.5 & 7.84 & 0 \\
\hline
\end{tabular}
\end{table}
\section{$\karma+$: Beyond uniform redistribution of karma}
\label{sec:non-uniform-redistribution}
A policy maker has  a considerable degree of freedom in the design of the karma payment and redistribution rules.
Thus far we have considered a simple scheme with uniform karma redistribution.
In this section, we briefly explore first steps towards the design of a karma redistribution scheme that further reduces congestion on the slow lane.
Namely, we conjecture that by redistributing more karma to commuters departing at the early/late stages of the congested period, we can reshape the departure patterns on the slow lane.
This new scheme is referred to as $\karma+$ and presented below. 
Let $n[t](d,\pi) = \sum_{b \in \Nat} \nu[t,b](d,\pi)$ be the mass of commuters departing at time $t$, and $r[t](d,\pi)$ be the karma redistribution of those commuters.
Motivated by the above intuition, we consider a family of piece-wise linear redistribution schemes that are parametrized by the tuple $(\tminre, \sigma)$.
The first parameter $\tminre \in \T$ specifies the departure time that receives the minimum redistribution.
The second parameter $\sigma > 1$ specifies a multiplier of the maximum redistribution karma with respect to $\tminre$ at the first and last departure times, i.e., $r[t^1](d,\pi) = r[t^T](d,\pi) = \sigma \: r[\tminre](d,\pi)$. 
The redistribution at other departure times are then derived via a linear interpolation given by
\begin{align}
    \label{eq:time-redistribution}
    r[t](d,\pi) = m[t] \: r[\tminre](d,\pi), \quad m[t] = \begin{cases}
        \frac{\sigma \: (\tminre - t) + t - t^1}{\tminre - t^1}, & t \leq \tminre, \\
        \frac{\sigma \: (t - \tminre) + t^T - t}{t^T - \tminre} & t > \tminre.
    \end{cases}
\end{align}
An example is given in the right panel of Figure~\ref{fig:results-non-uniform-redistribution}b.
For the preservation of karma, the total karma redistribution must equal the total karma payment, which is equivalent to the following:
\begin{align}
\label{eq:time-redistribution-karma-preservation}
\sum_t n[t](d,\pi) \: r[t](d,\pi) = \pbar(d,\pi).
\end{align}
The following proposition shows that the redistributed karma corresponding to each departure time can be uniquely determined once we know the departure pattern and the average karma payment. 
\begin{proposition}\label{prop:unique-redistribution}
    Given the mass of commuters departing at each time $n[t](d,\pi)$, for all $ t\in \T$ and the average karma payment $\pbar(d,\pi)$, the redistribution vector $r[t](d,\pi)$ is unique and continuous in $(d,\pi)$. 
\end{proposition}
\begin{proof}
The proof is included in Appendix~\ref{proof:unique-redistribution}.
\end{proof}





Accordingly, we can derive the karma transition probabilities similar to Section~\ref{sec:karma-transition} as follows:
\begin{equation}
    \Prob[k^+ \mid k,t,b,o](d,\pi) 
    = \begin{cases}
        f[t](d,\pi), & o = \textup{fast}, \text{ and } k^+ = k - b + \ceil{r[t](d,\pi)}, \\
        1-f[t](d,\pi), & o =  \textup{fast}, \text{ and } k^+ = k - b + \floor{r[t](d,\pi)}, \\
        f[t](d,\pi), & o =  \textup{slow}, \text{ and } k^+ = k + \ceil{r[t](d,\pi)}, \\
        1-f[t](d,\pi), & o =  \textup{slow}, \text{ and } k^+ = k + \floor{r[t](d,\pi)}, \\
        0, &\text{otherwise},
    \end{cases}
\end{equation}
where $f[t](d,\pi) = r[t](d,\pi) - \floor{r[t](d,\pi)}$.

It is then straightforward to verify that the corresponding karma transition function $\kappa[k^+ \mid k, t, b](d,\pi)$ satisfies Assumption~\ref{as:karma-preservation}, which is guaranteed by constraint~\eqref{eq:time-redistribution-karma-preservation}.
Therefore, Proposition~\ref{prop:sne-existence} can be readily extended to guarantee the existence of a \gls{SNE} under $\karma+$.


\begin{minipage}[t]{0.45\textwidth}
    \vspace{0pt}
    \centering
    \footnotesize
    
    \def \tleft {33.42651pt}
    \def \tright {21.73213pt}
    \def \tleftt {15.36607pt}
    \def \trightt {5.64877pt}
    \includegraphics[trim={\tleft{} 0 \tright{} 0}]{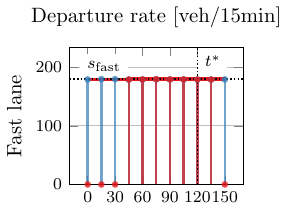}
    \hfill
    \includegraphics[trim={\tleftt{} 0 \trightt{} 0}]{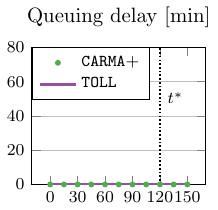}

    \def \tright {25.11082pt}
    \def \trightt {7.82932pt}
    \includegraphics[trim={\tleft{} 0 \tright{} 0}]{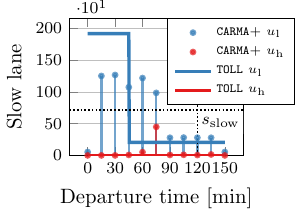}
    \hfill
    \includegraphics[trim={\tleftt{} 0 \trightt{} 0}]{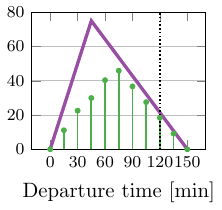}
    
    (a)
    \captionof{figure}{Results for time-dependant karma redistribution $\karma+$.
        (a) Left: departure rate in $\karma+$ versus $\toll$.
        Right: queuing delay over time.
        (b) Left: threshold bid $b^*[t]$. Right: karma redistribution $r[t]$.
        (c) Performance comparison of $\karma+$ versus $\toll$ and $\nom$.}
    \label{fig:results-non-uniform-redistribution}
\end{minipage}
\hspace{1mm}
\vline
\hspace{1mm}
\begin{minipage}[t]{0.45\textwidth}
    \vspace{0pt}
    \raggedleft
    \footnotesize
	
    \def \tleft {15.36607pt}
    \def \tright {10.00989pt}
    \includegraphics[trim={\tleft{} 0 \tright{} 0}]{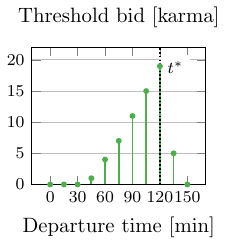}
    \hspace{9mm}
    \def \tleftt {11.11601pt}
    \def \trightt {10.91267pt}
    \includegraphics[trim={\tleft{} 0 \tright{} 0}]{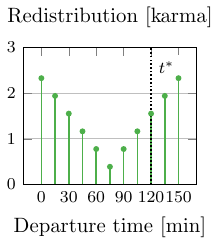}

    \centering
    (b)
    \medskip
    \hrule
    \smallskip
    \raggedleft
	
    \def \tleft {23.72046pt}
    \def \tright {23.72046pt}
    \def \tleftt {15.36607pt}
    \def \trightt {3.48428pt}
    \includegraphics[trim={\tleft{} 0 \tright{} 0}]{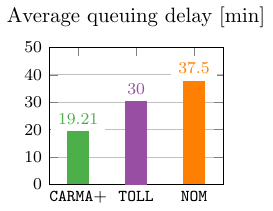}
    \hspace{9mm}
    \includegraphics[trim={\tleftt{} 0 \trightt{} 0}]{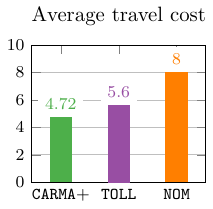}
	
    \centering
    (c)
\end{minipage}
\bigskip

In Figure~\ref{fig:results-non-uniform-redistribution}, we consider the same setting with homogeneous commuters as in Section~\ref{sec:case1} but use the time-varying karma redistribution scheme introduced above.
As shown in the right panel of Figure~\ref{fig:results-non-uniform-redistribution}b, we redistribute the least amount of karma to $\tminre=75$min (the center of the congested period) and give $\sigma=4$ times as much karma to the earliest and latest departure times.
Remarkably, this heuristic scheme leads to a significant congestion reduction on the slow lane; see the reduced departure rates and queuing delays in Figure~\ref{fig:results-non-uniform-redistribution}a.
As shown in Figure~\ref{fig:results-non-uniform-redistribution}c, compared to $\toll$, we can further decrease the average queuing delay by 36\% and the average travel cost by 16\%. 
These findings unfold the great potential of karma schemes for traffic demand management:
With a simple mechanism (e.g., uniform karma redistribution), it achieves a comparable efficiency as existing optimal monetary schemes in a more equitable manner, while given a better design, it can even surpass by indirectly affecting the departure pattern on the slow lane.
The optimal design of $\karma+$, as well as its sensitivity to private information such as the \gls{VOT} process, are open problems for future research.
\section{Conclusions}
\label{sec:conclusion}
This study proposes $\karma$ as a non-monetary solution to the bottleneck congestion in the morning commute. 
Under $\karma$, each commuter is endowed with non-monetary credits called karma used to bid for access to a noncongested fast lane.
The resulting commuter behaviors are modelled as a \acrfull{DPG} and the existence of a \acrfull{SNE} is proved.
This dynamic model enables the introduction of a novel kind of heterogeneity in the \acrfull{VOT}, namely, heterogeneity in how the \gls{VOT} varies dynamically on each day.
It also enables the study of the long-term fairness of $\karma$.
It is demonstrated, both analytically and numerically, that $\karma$ is equitable with respect to income and leads to a strong Pareto improvement for all commuter types with respect to no policy intervention.
These are properties that fail to hold in existing monetary schemes.

Through numerical analysis, it is shown that the fairness properties of $\karma$ do not come at a cost of reduced efficiency, as $\karma$ leads to similar departure and congestion patterns as an optimal tolling scheme and closely approaches the system optimum (when only the fast lane can be regulated).
It is also shown that $\karma$ is robust to heterogeneity in the desired arrival time and can adapt to this heterogeneity in a decentralized and privacy-preserving manner.
Finally, it is demonstrated that a mechanism with more advanced karma redistribution, called $\karma+$, further reduces congestion and thus obtains a higher efficiency than the optimal tolling on the fast lane.

There are several directions to extend the current work. 
First, it remains open to extending the strong Pareto improvement guarantee of $\karma$ to more general karma payment and redistribution schemes, as well as more classes of heterogeneity (e.g., the heterogeneity in desired arrival time and delay penalties). 
Second, observing the promising performance of a heuristic non-uniform karma redistribution, we believe the optimal design of karma payment and redistribution scheme is an exciting future direction.  
Meanwhile, the optimal design will likely require information about the distribution of desired arrival times and other private information. It is thus also interesting to investigate the trade-off between efficiency and privacy. 
Finally, we believe the application of karma in transportation systems goes beyond the morning commute problem and hope this study could inspire a series of future works on demand management with karma. 


\section*{Acknowledgments}
The authors want to thank Angel Manzano Carnicer for the fruitful and interesting discussions.
Research supported by NCCR Automation, a National Centre of
Competence in Research, funded by the Swiss National Science
Foundation (grant number $180545$).

\newpage
\appendix
\section{Bottleneck model preliminaries}\label{sec:bottleneck}
In this section, we review some results from the classical bottleneck model, following the same notations introduced in $\karma$.  

Consider a population of $N$ commuters passing through a single bottleneck with capacity $s$. Under the optimal tolling scheme, the bottleneck is split into two lanes: a tolled lane with capacity $\sfast$ and a regular lane with capacity $\sslow$. 
Following the setting of $\karma$, we assume commuters are heterogeneous in both VOT and desired arrival time, and they can be classified into types. Commuters in the same type $\tau\in\Gamma$ share the same desired arrival time $t^*_\tau$ but may differ in their VOT values (due to the dynamic \gls{VOT} process). Nevertheless, the distribution of VOT values within each type is stationary and exogenous. Besides, all commuters share the same penalty ratios as per Assumption~\ref{as:constant_ratio_penalties}.

Hence, the \emph{travel cost} for commuters of type $\tau$ with \gls{VOT} value $u$ who enter the fast lane at time $t$ reads
\begin{align}
\label{eq:cost_fast_lane_bn}
    \cfast_ \tau[u](t) = \begin{cases}
    u \: \beta \: (t^*_\tau-t), & t\leq t^*_ \tau, \\
     u \: \gamma \: (t-t^*_ \tau), & t>t^*_ \tau.
  \end{cases}
\end{align} 
and their total cost is given by $\hat c^\textup{fast}_\tau[u](t)\coloneqq c^\textup{fast}_\tau[u](t)+p(t)$, where $p(t)$ is the toll price at time $t$.

For those traveling on the slow lane, the travel cost, as well as the total travel cost, is 
\begin{align}
\label{eq:cost_slow_lane_bn}
 \cslow_ \tau[u](t) = \begin{cases}
     u \left[\alpha \: \frac{q(t)}{\sslow} + \beta \: (t^*_ \tau - t - \frac{q(t)}{\sslow}) \right], & t + \frac{q(t)}{\sslow}\leq t^*_\tau, \\
     u \left[\alpha \: \frac{q(t)}{\sslow} + \gamma \: (t + \frac{q(t)}{\sslow} - t^*_ \tau) \right], & t + \frac{q(t)}{\sslow}>t^*_ \tau,
 \end{cases}
\end{align}
where $q(t)$ denotes the queue length on the slow lane at time $t$.

The objective of commuters is to minimize their own total travel cost by choosing the lane and departure time. For commuters of type $\tau$ and traveling on the slow lane, the equilibrium condition yields the derivative of queue length on the slow lane with respect to departure time $t$ during the congestion period as follows:
\begin{align}\label{eq:eq-queue-derv}
     q'(t) = 
    \begin{cases}
        \frac{\beta}{\alpha-\beta} \: \sslow, & t + \frac{q(t)}{\sslow}\leq t^*_\tau, \\
        -\frac{\gamma}{\alpha+\gamma}\sslow, & t + \frac{q(t)}{\sslow}>t^*_\tau.
    \end{cases}
\end{align}
Similarly, the derivative of the optimal toll price for the departure window of commuters of type $\tau$ and with urgency $u$  on the fast lane is given by
\begin{align}\label{eq:eq-toll-derv}
    p'(t) = 
    \begin{cases}
        u\beta, & t<t^*_\tau,\\
        -u\gamma, & t>t^*_\tau.
    \end{cases}
\end{align}
Note that \eqref{eq:eq-queue-derv} applies to $\nom$ by replacing $\sslow$ with $s$. These results are used to derive the equilibrium cost and the optimal toll price presented below.

\subsection{Homogeneous desired arrival time}
\label{sec:bottleneck-homo-t*}
In this section, we present the results in the case of homogeneous desired arrival time, i.e., $t^*_\tau \equiv t^*,\;\textup{for all } \tau\in\Gamma$. 

Using \eqref{eq:eq-queue-derv}, we can easily draw the aggregate departure pattern under $\nom$ and derive the start and end of the congestion period ($\tstart$ and $\tend$), as well as the time with maximum queuing delay ($\tpeak$), as follows:
\begin{align}
    \tstart = t^* - c^*/\beta,\quad \tend = t^*+c^*/\gamma, \quad \tpeak = t^*-c^*/\alpha, \label{eq:t-start-t-end-t-peak}
\end{align}
where
\begin{align}
    c^*= \frac{\beta\gamma}{\beta+\gamma}\frac{N}{s} \label{eq:c-star}
\end{align}
is the \gls{VOT}-normalized equilibrium cost. For commuters with \gls{VOT} $u$, the equilibrium cost is given by $c^*[u]=uc^*$. 
Another well-known result is that the queuing delay takes half of the total cost~\cite{arnott1990economics} and thus $\bar{t}^q[u]=\frac{c^*}{2\alpha},\;\textup{for all } u$. 
These lead to the performance measures listed in the first column of Table~\ref{tab:measure}. 

\begin{figure}[htb]
\begin{subfigure}{0.5\textwidth}
\centering
\includegraphics[width=\textwidth]{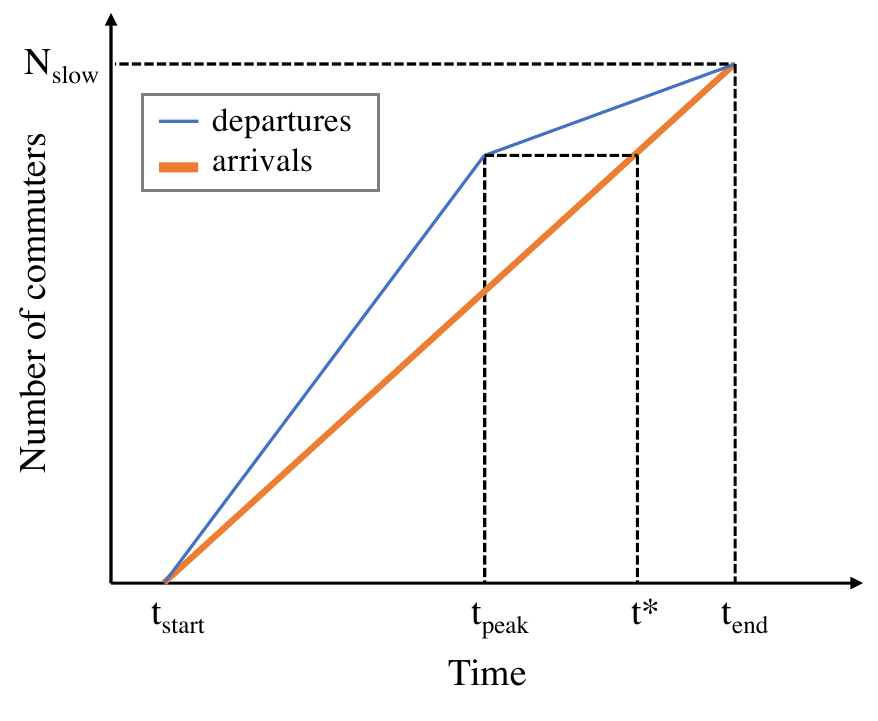}
\caption{Equilibrium departures/arrivals on slow lane.}
\label{fig:model_eq}
\end{subfigure}
\begin{subfigure}{0.5\textwidth}
\centering
\includegraphics[width=\textwidth]{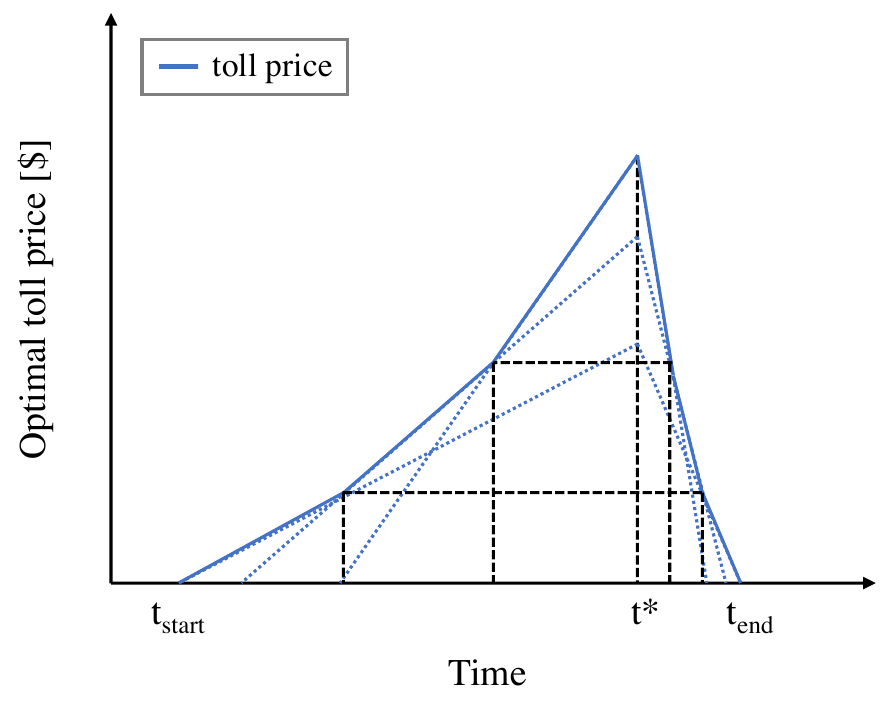}
\caption{Optimal toll price $p^*(t)$ on fast lane.}
\label{fig:model_toll}
\end{subfigure}
\caption{Illustration of equilibrium under $\toll$ with same desired arrival time but heterogeneous VOT.}
\label{fig:model}
\end{figure}

Under $\toll$, commuters with higher VOT $u$ would enter the fast lane and depart closer to $t^*$, and the departure windows are exclusive and allocated on both sides of $t^*$ \cite{xiao2013managing}. 
An illustration of the equilibrium result under $\toll$ is presented in Figure~\ref{fig:model}. 

Note that the congestion period under $\toll$ remains the same as $\nom$, but the system average queuing delay reduces to $\frac{\sslow}{s}\frac{c^*}{2\alpha}$ because the queue only emerges on the slow lane.
Let $\hat{u}$ be the threshold \gls{VOT} such that commuters with $u>\hat{u}$ all enter the fast lane and those with $u<\hat{u}$ all enter the slow lane. Then, the fraction of commuters with $\hat{u}$ who enter the slow lane is given by 
\begin{align}
    \hat{r} = \frac{1}{\Prob[\hat{u}]}\left(\frac{s_\textup{slow}}{s} - \sum_{u<\hat{u}}\Prob[u]\right),
\end{align}
where $\Prob[u]$ gives the fraction of commuters with \gls{VOT} $u$. 

Accordingly, the average queuing delay and travel cost for commuters with \gls{VOT} $u$ are derived as 
\begin{align}
    \bar{t}^q[u] &= 
    \begin{cases}
    \frac{c^*}{2\alpha}, & u<\hat{u}, \\ 
    \hat{r} \: \frac{c^*}{2\alpha}, & u=\hat{u},\\
    0, & u>\hat{u},
    \end{cases} \label{eq:eq-queue-homo-t*}\\
    c^*[u] &= 
    \begin{cases} 
    u \: c^*, & u<\hat{u}, \\ 
    u \: c^*\left[1 - (1-\hat{r}) \: \frac{s}{2 \: \sfast} \: \Prob[\hat{u}]\right], & u=\hat{u},\\
    u \: c^*\left(1-\frac{s}{2 \: \sfast} \: \Prob[u]\right), & u>\hat{u}.
    \end{cases}\label{eq:eq-cost-homo-t*}
\end{align}
\eqref{eq:eq-queue-homo-t*} and \eqref{eq:eq-cost-homo-t*} yield the formulae presented in the second column of Table~\ref{tab:measure}.

\subsection{Heterogeneous desired arrival time}
\label{sec:bottleneck-hetero-t*}
\begin{figure}[H]
\begin{subfigure}{0.95\textwidth}
\centering
\includegraphics[width=\textwidth]{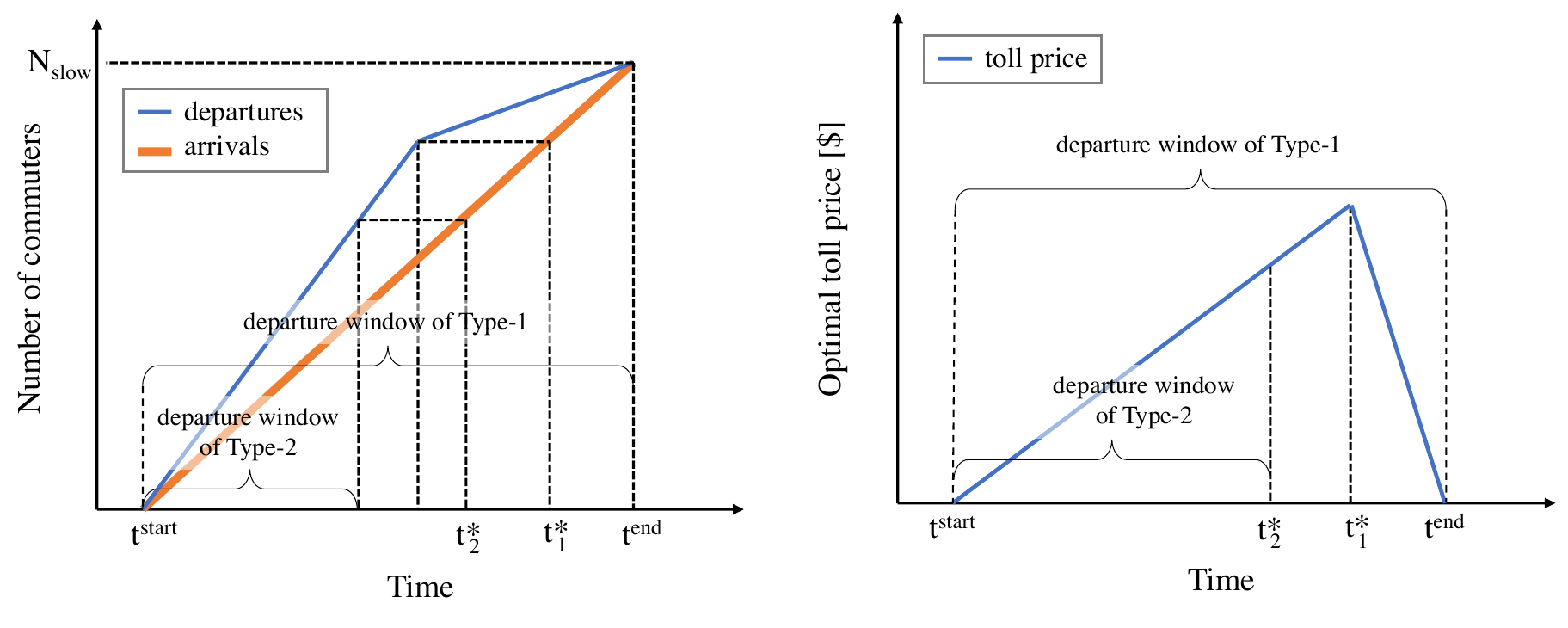}
\caption{Type 2 commuters have early desired arrival time and smaller group size.}
\end{subfigure}
\begin{subfigure}{0.95\textwidth}
\centering
\includegraphics[width=\textwidth]{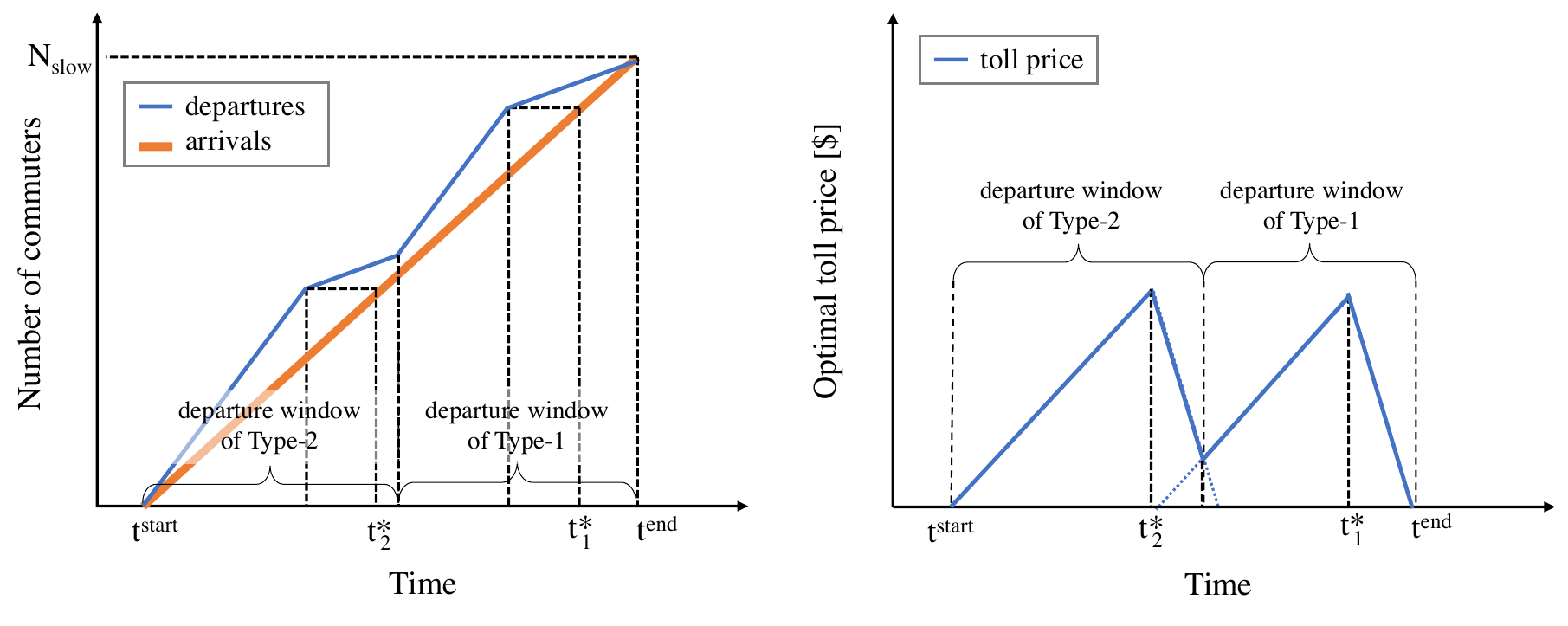}
\caption{Type 2 commuters have early desired arrival time and similar group size.}
\end{subfigure}
\caption{Illustration of equilibrium under $\toll$ with heterogeneous VOT and desired arrival time (left: departures/arrivals on slow lane; right: optimal toll price on fast lane).}
\label{fig:model-hetero-t*}
\end{figure}

We proceed to specify the results for heterogeneous desired arrival times with two types, namely, Type 1 and Type 2 commuters. For notation simplicity, we denote their desired arrival times as $t^*_1$ and $t^*_2$, respectively. 
As suggested by \cite{arnott1988schedule}, the bottleneck may exhibit a single peak (e.g., Figure~\ref{fig:model-hetero-t*}a) or multiple peaks (e.g., Figures~\ref{fig:model-hetero-t*}c) on the slow lane.
Due to \eqref{eq:eq-queue-derv}, a single peak implies that, if both types of commuters use the slow lane, one of them must depart on one side of its desired arrival time. The departure windows on the fast lane, however, depend on both the desired arrival time and the VOT value of each commuter. 

Figure~\ref{fig:model-hetero-t*} illustrates a special case where commuter types share the same set of VOT values and all commuters using the fast lane share the same VOT value. In Figure~\ref{fig:model-hetero-t*}a, there are much fewer Type 2 commuters than Type 1 and $t^*_2<t^*_1$. Hence, all Type 2 commuters depart before $t^*_2$, and both the departure pattern on the slow and the toll price on the fast lane exhibit a single peak. Differently, the number Type 2 commuters in Figure~\ref{fig:model-hetero-t*}b is comparable to Type 1. Hence, in this case, there are two peaks in the slow-lane departure curve and the fast-lane toll price. 

Figure~\ref{fig:model-hetero-t*} also corresponds to the first two cases specified in Section~\ref{sec:case4}. The third case (i.e., Type 2 commuters are rather few and have a much earlier desired arrival time) is trivial and the plot looks almost the same as Figure~\ref{fig:model}. In what follows, we derive the closed-form performance measures for the three cases.

\subsubsection{Single peak}
When the congestion period exhibits a single peak, the aggregate departure pattern remains the same as in Section~\ref{sec:bottleneck-homo-t*}. 
Hence, the system average queuing delay is still $\frac{c^*}{2\alpha}$ under $\nom$ and $\frac{\sslow}{s}\frac{c^*}{2\alpha}$ under $\toll$. 
Besides, the last Type 2 commuter under both $\nom$ and $\toll$ arrives at $t^*_2$. 
Using this result, we can easily derive the departure window of Type 2 as
\begin{align}
    \tstart_2 = \tstart,\quad \tend_2 = \frac{\alpha-\beta}{\alpha} \: t^*_2 + \frac{\beta}{\alpha} \: \tstart.
\end{align}
Under $\nom$, the normalized travel costs of the two types are 
\begin{align}
    c^*_1 &= c^*,\label{eq:eq-cost-hetero-t*-single-type1}\\
    c^*_2 &= c^* - \beta \: (t^*_1 - t^*_2),\label{eq:eq-cost-hetero-t*-single-type2}
\end{align}
and their average queuing delays are 
\begin{align}
    \bar{t}^q_1 &= \frac{c^*}{2 \: \alpha} + \frac{N_2}{N_1} \: \frac{\beta}{2 \: \alpha} \: (t^*_1 - t^*_2),\label{eq:eq-queue-hetero-t*-single-type1}\\
    \bar{t}^q_2 &= \frac{c^*}{2 \: \alpha} - \frac{\beta}{2 \: \alpha} \: (t^*_1 - t^*_2).\label{eq:eq-queue-hetero-t*-single-type2}
\end{align}
Accordingly, the system normalized average travel cost is 
\begin{align}\label{eq:eq-avg-cost-hetero-t*-single}
    \cbarnorm = c^* - \frac{N_2}{N} \: \beta \: (t^*_1 - t^*_2).
\end{align}

Since commuters share the same \gls{VOT} process, the overall departure pattern with respect to VOT values remains the same as in Section~\ref{sec:bottleneck-homo-t*}. Therefore, the average queuing delay and travel cost follow the same forms as \eqref{eq:eq-queue-homo-t*} and \eqref{eq:eq-cost-homo-t*}. For commuters of type $\tau$, they are 
\begin{align}
    \bar{t}^q_\tau[u] &= 
    \begin{cases}
    \bar{t}^q_\tau, & u<\hat{u}, \\ 
    \hat{r} \: \bar{t}^q_\tau, & u=\hat{u},\\
    0, & u>\hat{u},
    \end{cases} \label{eq:eq-queue-hetero-t*-single}\\
    c^*_\tau[u] &= 
    \begin{cases} 
    u \: c^*_\tau, & u<\hat{u}, \\ 
    u \: c^*_\tau\left[1 - (1-\hat{r}) \: \frac{s}{2 \: \sfast} \: \Prob[\hat{u}]\right], & u=\hat{u},\\
    u \: c^*_\tau\left(1-\frac{s}{2\: \sfast} \: \Prob[u]\right), & u>\hat{u},
    \end{cases}\label{eq:eq-cost-hetero-t*-single}
\end{align}
where $c^*_\tau$ and $\bar{t}^q_\tau$ take the values in \eqref{eq:eq-cost-hetero-t*-single-type1}--\eqref{eq:eq-queue-hetero-t*-single-type2}.

\subsubsection{Double peak}
When there are two peaks, the two commuter types have two exclusive departure windows. Let $\tbnd$ denote the boundary arrival time (the arrival time of the last Type $2$ commuter and the first Type $1$ commuter). 
Note that \eqref{eq:eq-queue-derv} can also be written as a function of arrival time $\tarr = t + \frac{q(t)}{s}$ \cite{newell1987morning}, i.e., 
\begin{align}
    q'(\tarr) = 
    \begin{cases}
        \frac{\beta}{\alpha} \: s, & \tarr \leq t^*_\tau,\\
        -\frac{\gamma}{\alpha} \: s, & \tarr > t^*_\tau.
    \end{cases}
\end{align}
Then, the starting, ending, and boundary arrival times satisfy the following conditions:
\begin{align}\label{eq:eq-condition-hetero-t*-double}
    \begin{cases}
        (\tend - \tstart) \: s = N,\\
        (\tbnd - \tstart) \: s = N_2,\\
        (\tend - \tbnd) \: s = N_1,\\
        \beta \: (t^*_2 - \tstart) - \gamma \: (\tbnd - t^*_2) = \gamma \: (\tend - t^*_1) - \beta \: (t^*_1 - \tbnd). 
    \end{cases}
\end{align}
The last equation in \eqref{eq:eq-condition-hetero-t*-double} states that the queue length observed by the last Type 2 commuter is the same as that observed by the first Type 1 commuter. 
Solving the system of equations \eqref{eq:eq-condition-hetero-t*-double} yields
\begin{align}
    \tstart &= \frac{1}{2}\left(t^*_1+t^*_2  - \frac{c^*}{\beta} - \frac{N_2}{s}\right),\\
    \tend &= \frac{1}{2}\left(t^*_1+t^*_2 + \frac{c^*}{\gamma} + \frac{N_1}{s}\right),\\
    \tbnd &= \frac{1}{2}\left(t^*_1+t^*_2 - \frac{\gamma}{\beta+\gamma} \: \frac{N_1}{s} + \frac{\beta}{\beta+\gamma} \: \frac{N_2}{s}\right).
\end{align}

Accordingly, the normalized travel cost of the two groups are 
\begin{align}
    c^*_1 &= \frac{1}{2}\left[c^* + \gamma \: (t^*_2 - t^*_1 + \frac{N_1}{s})\right],\label{eq:eq-cost-hetero-t*-double-type1}\\
    c^*_2 &= \frac{1}{2}\left[c^* + \beta \: (t^*_2 - t^*_1 + \frac{N_2}{s})\right], \label{eq:eq-cost-hetero-t*-double-type2}
\end{align}
and the system normalized average travel cost is
\begin{align}\label{eq:eq-avg-cost-hetero-t*-double}
    \cbarnorm = \frac{1}{2} \left[ c^* + \frac{\gamma \: N_1 + \beta \: N_2}{N} \: (t^*_2 - t^*_1) + \frac{\gamma \: N^2_1 + \beta N^2_2}{N \: s}\right].
\end{align}
With some algebra, we derive the average queuing delay for each commuter type as
\begin{align}
    \bar{t}^q_1 =& \frac{c^*}{2 \: \alpha} + \frac{\gamma}{2 \: \alpha}\left(t^*_2 - t^*_1 + \frac{N_1}{s}\right)\label{eq:eq-queue-hetero-t*-double-type1}\\
    & - \frac{s}{8 \: \alpha \: N_1}\left[\beta \: \left(t^*_2 - t^*_1 + \frac{c^*}{\gamma} - \frac{N_1}{s}\right)^2 + \gamma \: \left(t^*_2 - t^*_1 + \frac{c^*}{\gamma} + \frac{N_1}{s}\right)^2\right],\nonumber\\
    \bar{t}^q_2 =& \frac{c^*}{2 \: \alpha} + \frac{\beta}{2 \: \alpha}\left(t^*_2 - t^*_1 + \frac{N_2}{s}\right)\label{eq:eq-queue-hetero-t*-double-type2}\\
    & - \frac{s}{8 \: \alpha \: N_2}\left[\beta \: \left(t^*_2 - t^*_1 + \frac{c^*}{\beta} + \frac{N_2}{s}\right)^2 + \gamma \: \left(t^*_2 - t^*_1 + \frac{c^*}{\beta} - \frac{N_2}{s}\right)^2\right].\nonumber
\end{align}
Finally, the system-level average queuing delay is
\begin{align}\label{eq:eq-avg-queue-hetero-t-double}
    \bar{t}^q = &\frac{c^*}{2 \: \alpha} + \frac{\gamma \: N^2_1 + \beta \: N^2_2}{2 \: \alpha \: N \: s}- \frac{(\beta+\gamma) \: s}{4 \: \alpha \: N}(t^*_2 - t^*_1)^2 \\
    &- \frac{s}{8 \: \alpha \: N}\left\{\beta\left[\left(\frac{c^*}{\gamma} - \frac{N_1}{s}\right)^2 + \left(\frac{c^*}{\beta} + \frac{N_2}{s}\right)^2 \right]
    + \gamma\left[\left(\frac{c^*}{\gamma} + \frac{N_1}{s}\right)^2 + \left(\frac{c^*}{\beta} - \frac{N_2}{s}\right)^2 \right]\right\}\nonumber.
\end{align}

Similar to the single-peak scenario, the performance measures under $\toll$ are in line with the forms \eqref{eq:eq-queue-homo-t*} and \eqref{eq:eq-cost-homo-t*} but differ in the type-specific values of $\bar{t}^q_\tau$ and $c^*_\tau$. Hence, they share the same formulae of \eqref{eq:eq-queue-hetero-t*-single} and \eqref{eq:eq-cost-hetero-t*-single} though with type-specific travel cost and average queuing delay given in \eqref{eq:eq-cost-hetero-t*-double-type1}--\eqref{eq:eq-queue-hetero-t*-double-type2}. The system average queuing delay is \eqref{eq:eq-avg-queue-hetero-t-double} multiplied by a factor of $\sslow/s$.

\subsubsection{Precongestion}
In this scenario, the two commuter types have disjoint departure windows. Since the number of Type 2 commuters is below the bottleneck capacity as well as the slow lane capacity, all commuters of that type can pass the bottleneck at $t^*_2$ without enduring any congestion. Type 1 commuters then start departing and generate bottleneck congestion independently. There, the performance measures can be easily derived as reported in Table~\ref{tab:measure-hetero-t*}.

\begin{table}[H]
\caption{Performance measures with heterogeneous desired arrival time.}
\label{tab:measure-hetero-t*}
\centering
\footnotesize
\renewcommand{\arraystretch}{1.5}
\begin{tabular}{|p{0.34\textwidth}p{0.02\textwidth}|p{0.31\textwidth}|p{0.2\textwidth}|}
\hline
\textbf{Name}     &  & \textbf{Benchmark}  & \textbf{Optimal tolling}\\
& & (``$\nom$'') &  (``$\toll$'') \\
\hline
\hline
\multicolumn{4}{|l|}{{\textbf{Single peak}}}\\
\hline
System average queuing delay     & $\bar{t}^q$      & $\frac{c^*}{2\alpha}$           &      $\frac{\sslow}{s} \: \frac{c^*}{2\alpha}$     \\
\hline
System average travel cost       & $\cbarnorm$        & $c^* - \frac{N_2}{N}\beta(t^*_1 - t^*_2)$ &                  $\sum_\tau g_\tau \: \cbarnorm_\tau$      \\
\hline
Type average queuing delay & $\bar{t}^q_1$ &    $\frac{c^*}{2\alpha} + \frac{N_2}{N_1}\frac{\beta}{2\alpha}(t^*_1 - t^*_2)$              &    $\sum_u \Prob_\tau[u] \: \bar{t}^q_\tau[u]$   \\
& $\bar{t}^q_2$ & $\frac{c^*}{2\alpha} - \frac{\beta}{2\alpha}(t^*_1 - t^*_2)$ &  with \eqref{eq:eq-queue-hetero-t*-single}\\
\hline
Type average travel cost & $\cbarnorm_1$   &         $c^*$              &    $\frac{1}{\ubar_\tau}\sum_u \Prob_\tau[u] \: c^*_\tau[u]$\\
& $\cbarnorm_2$   & $c^* - \beta(t^*_1 - t^*_2)$  &  with \eqref{eq:eq-cost-hetero-t*-single}\\
\hline
\hline
\multicolumn{4}{|l|}{{\textbf{Double peak}}}\\
\hline
System average queuing delay     & $\bar{t}^q$      &   \eqref{eq:eq-avg-queue-hetero-t-double}      &      $\frac{\sslow}{s}$ \eqref{eq:eq-avg-queue-hetero-t-double}     \\
\hline
System average travel cost       & $\cbarnorm$        & \eqref{eq:eq-avg-cost-hetero-t*-double} &     $\sum_\tau g_\tau \: \cbarnorm_\tau$    \\
\hline
Type average queuing delay & $\bar{t}^q_1$ &   \eqref{eq:eq-queue-hetero-t*-double-type1}  &    $\sum_u \Prob_\tau[u] \: \bar{t}^q_\tau[u]$\\
& $\bar{t}^q_2$ &   \eqref{eq:eq-queue-hetero-t*-double-type2}   &  with \eqref{eq:eq-queue-hetero-t*-single}\\
\hline
Type average travel cost & $\cbarnorm_1$   &   $\frac{1}{2}\left[c^* + \gamma(t^*_2 - t^*_1 + \frac{N_1}{s})\right]$           &    $\frac{1}{\ubar_\tau} \sum_u \Prob_\tau[u] \: c^*_\tau[u]$\\
& $\cbarnorm_2$ & $\frac{1}{2}\left[c^* + \beta(t^*_2 - t^*_1 + \frac{N_2}{s})\right]$   &  with \eqref{eq:eq-cost-hetero-t*-single}\\
\hline
\hline
\multicolumn{4}{|l|}{{\textbf{Precongestion}}}\\
\hline
System average queuing delay     & $\bar{t}^q$      & $\frac{c^*}{2\alpha} (\frac{N_1}{N})^2$ &      $\frac{\sslow}{s} \: \frac{c^*}{2\alpha}(\frac{N_1}{N})^2$     \\
\hline
System average travel cost       & $\cbarnorm$        &  $c^* \: (\frac{N_1}{N})^2$   &     $\sum_\tau g_\tau \: \cbarnorm_\tau$    \\
\hline
Type average queuing delay & $\bar{t}^q_1$ & $\frac{c^*}{2\alpha}\frac{N_1}{N}$   &    $\sum_u \Prob_\tau[u] \: \bar{t}^q_\tau[u]$\\
& $\bar{t}^q_1$ & 0 &  with \eqref{eq:eq-queue-hetero-t*-single}\\
\hline
Type average travel cost & $\cbarnorm_1$   & $c^* \: \frac{N_1}{N}$ &    $\frac{1}{\ubar_\tau} \sum_u \Prob_\tau[u] \: c^*_\tau[u]$\\
& $\cbarnorm_2$ & 0 &  with \eqref{eq:eq-queue-hetero-t*-single}\\
\hline
\end{tabular}
\end{table}

\section{Karma mechanism: theoretical framework}
\label{sec:karma-preliminaries}
In this section, we revisit the \emph{karma mechanism} firstly introduced in~\cite{elokda2023self} that forms the foundation of the analysis carried out in Section~\ref{sec:karma-bottleneck}. In brief, karma is used to repeatedly allocate a resource to a group of competing agents, viz. commuters in \karma, over an infinite time horizon. 
At each time step, an agent endowed with an integer quantity $k \in \Nat$, called \emph{karma}, can submit an integer  karma bid $b \in \B[k]=\{0,\dots,k\}$.
Apart from the karma, each agent also features an urgency state $u_\tau \in \U=\{u_1,\dots,u_{M_\tau}\}$, $u>0$, viz. the \gls{VOT} in $\karma$, evolving according to an exogenous
irreducible Markov chain $\phi_\tau[u^+ \mid u]$ for every agent type $\tau \in \Gamma\subset \Nat$.

Formally, the karma mechanism is modeled as a \emph{\acrfull{DPG}}~\cite{elokda2021dynamic}. Namely, the number of agents $N$ is assumed large, thus they can be approximated by a continuum of mass.
The distribution of agent types is compactly denoted by $g\in\Delta(\Gamma)$, where $g_\tau\in[0,1]$ is the mass of agents in type $\tau\in\Gamma$.
Accordingly, the time-varying joint \emph{type-state distribution} is given by
\begin{align}
    d \in \D = \left\{d \in \Real_+^{|\Gamma| \times |\X|} \left\lvert \; \sum_{u\in\U_\tau}\sum_{k\in\Nat} d_\tau[u,k] = g_\tau, \; \textup{for all } \tau \in\Gamma \right. \right\},
\end{align}
where $d_\tau[u,k]$ denotes the mass of agents in the \emph{static type} $\tau$ and \emph{dynamic state} $[u,k] \in \X_\tau$, and $\X_\tau = \U_\tau \times \Nat$.

At each time step, each agent can choose an action $a$ from a finite state-dependent discrete set $\A[u,k]$. The action includes the agent's bid $b$ as well as other decisions, viz. the departure time in \karma.
Agents of the same type $\tau$ follow the homogeneous randomized policy 
$\pi_\tau : \X_\tau \rightarrow \Delta(\A[u,k])$,
where $\pi_\tau[a \mid u,k]$ denotes the probabilistic weight that these agents place on action $a$ when in state $[u,k]$.
The concatenation of the policies of all types 
$\pi = (\pi_\tau)_{\tau\in\Gamma}$ is simply referred to as the \emph{policy}, and the space of policies is denoted by $\Pi$.

\begin{figure}[ht]
\centering
\includegraphics[width=\textwidth]{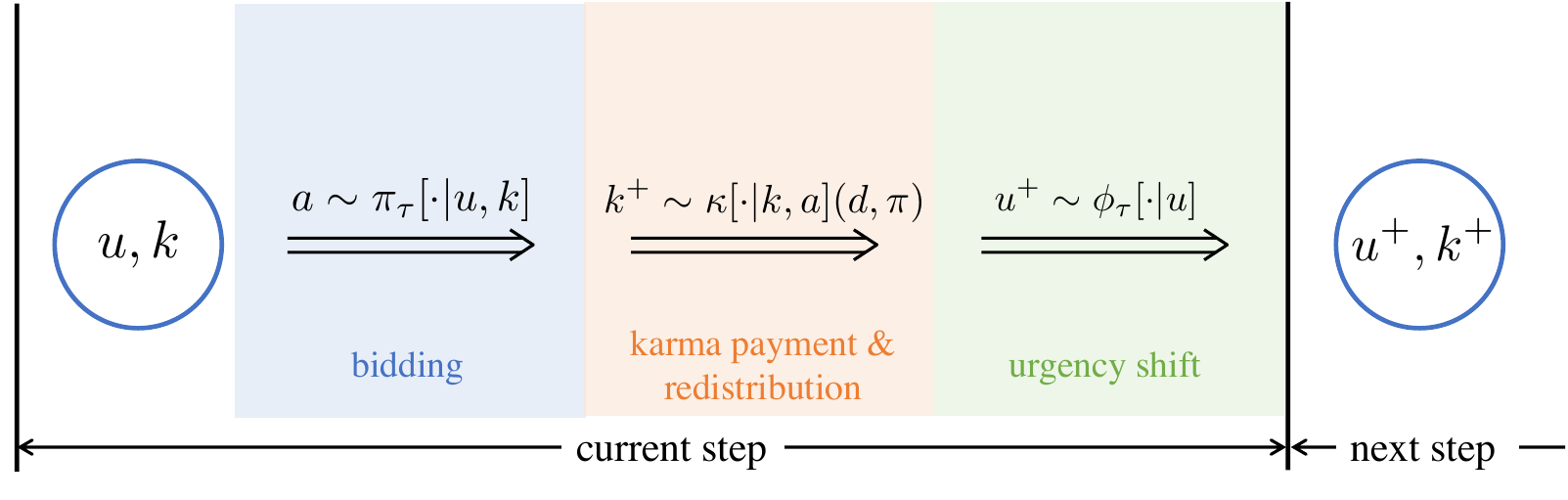}
\caption{One-step state transition of an agent of type $\tau$.}
\label{fig:karma_transit}
\end{figure}

In the \gls{DPG}, the tuple of type-state distribution and policy $(d,\pi)$ is referred to as the \emph{social state} because it gives a macroscopic description of the distribution of agents in the population as well as their behaviors. Hence, all the outcomes at each time step are functions of $(d,\pi)$. 
Let $\kappa[k^+ \mid k,a](d,\pi)$ be the \emph{karma transition function} that describes how the agent's karma changes between two consecutive time steps given its current karma $k$ and action $a$. Then, together with the urgency transition function $\phi_\tau[u^+ \mid u]$, the joint \emph{state transition function} is given by
\begin{align}
    \label{eq:state-transition}
    \rho_\tau[u^+, k^+ \mid u,k,a](d,\pi) = \phi_\tau[u^+ \mid u] \: \kappa[k^+ \mid k, a](d,\pi).
\end{align}
Figure~\ref{fig:karma_transit} illustrates a one-step state transition of an agent of type $\tau$.
Moreover, we define $\zeta_\tau[u,a](d,\pi)$ as the \emph{immediate reward function} of each type $\tau$ agent in urgency $u$ taking action $a$. Both the immediate reward and the karma transition are specified in Section~\ref{sec:karma-bottleneck}. Yet, two conditions are required to ensure that the karma mechanism is well defined.
\begin{assumption}[Continuity]
\label{as:continuity}
The immediate reward function $\zeta_\tau[u,a](d,\pi)$ and the 
karma transition function $\kappa[k^+ \mid k,a](d,\pi)$ are continuous in the social state $(d,\pi)$.
\end{assumption}
\begin{assumption}[Karma preservation in expectation]
\label{as:karma-preservation}
Karma is preserved in expectation for all $(d,\pi)$, i.e., $\E[k^+] = \E[k]$,
which expands to
\begin{multline}    
    \sum_{\tau \in \Gamma} \sum_{u \in \U_\tau} \sum_{k \in \Nat} d_\tau[u,k] \sum_{a \in \A[u,k]} \pi_\tau[a \mid u,k] \sum_{k^+ \in \Nat} \kappa[k^+ \mid k,a](d,\pi) \: k^+ \\
    = \sum_{\tau \in \Gamma} \sum_{u \in \U_\tau} \sum_{k \in \Nat} d_\tau[u,k] \: k.  \label{eq:karma-preservation}
\end{multline}
\end{assumption}
The readers are referred to~\cite{elokda2023self} for an in depth discussion on the assumptions and functions introduced above.

Given the social state, each agent faces a \acrfull{MDP}. 
Specifically, the expected reward of the agents of type $\tau$ is given by 
\begin{align}
    R_\tau[u,k](d,\pi) = \sum_{a \in \A[u,k]} \pi_\tau[a \mid u,k] \: \zeta_\tau[u,a](d,\pi), \label{eq:exp-immediate-reward}
\end{align}
and the state transition follows 
\begin{align}
    P_\tau[u^+,k^+ \mid u,k](d,\pi) = \sum_{a \in \A[u,k]} \pi_\tau[a \mid u,k] \: \rho_\tau[u^+,k^+ \mid u,k,a](d,\pi).
\end{align}
Accordingly, the expected return in the infinite horizon, also known as the value function, is derived as 
\begin{equation}\label{eq:value-function}
    V_\tau[u,k](d,\pi) 
    = R_\tau[u,k](d,\pi) + \delta \sum_{u^+ \in \U_\tau} \sum_{k^+ \in \Nat} P_\tau[u^+,k^+ \mid u,k](d,\pi) \: V_\tau[u^+,k^+](d,\pi),
\end{equation}
where $\delta\in(0,1]$ is the future discount factor. A smaller $\delta$ models a more myopic agent behavior.
To describe the rational decision of each agent, we also need to define the state-action value function, also called $Q$-function, as
\begin{equation}
    \label{eq:Q-function}
    Q_\tau[u,k,a](d,\pi)
    = \zeta_\tau[u,a](d,\pi) + \delta \sum_{u^+ \in \U_\tau} \sum_{k^+ \in \Nat} \rho_\tau[u^+,k^+ \mid u,k,a](d,\pi) \: V_\tau[u^+,k^+](d,\pi).
\end{equation}
Then, to maximize the long-term return, each agent chooses a policy based on the \emph{best response} correspondence, given by 
\begin{multline}
    B_\tau[u,k](d,\pi) = \left\{\sigma \in \Delta(\A[u,k]) \left \lvert \; \textup{for all } \sigma' \in \Delta(\A[u,k]), \vphantom{\sum_a} \right. \right.\\
    \left. \sum_{a \in \A[u,k]} (\sigma[a] - \sigma'[a]) \: Q_\tau[u,k,a](d,\pi) \geq 0 \right\}.
\end{multline}
By definition, $B_\tau[u,k](d,\pi)$ gives the set of randomized individual actions that maximize the $Q$-function, viz. the long-term reward, in state $[u,k]$ for agents of type $\tau$. The above framework 
 models  the rational decision-making process that each agent undergoes to select their strategy.

We are finally ready to formally define the equilibrium concept that we consider  under a karma mechanism.
\begin{definition}[\acrfull{SNE}]
\label{def:SNE}
A \acrlong{SNE} is a social state $(d^*,\pi^*) \in \D \times \Pi$ such that, for all $[\tau,u,k] \in \Gamma \times \U_\tau \times \Nat$,
\begin{align}
    d^*_\tau[u,k] &= \sum_{u^- \in \U_\tau} \sum_{k^- \in \Nat} d^*_\tau[u^-,k^-] \: P_\tau[u,k \mid u^-,k^-](d^*,\pi^*), \label{eq:SNE-1} \\
    \pi^*_\tau[\cdot \mid u,k] &\in B_\tau[u,k](d^*,\pi^*). \label{eq:SNE-2} 
\end{align}
\end{definition}

We conclude this section by stating the conditions for a karma mechanism to guarantee the existence of a \gls{SNE}  
(see~\cite{elokda2023self} for the proof).
\begin{theorem}[Existence of \gls{SNE}, Th.~1 \cite{elokda2023self}]
\label{th:Nash-exists}
Let Assumption~\ref{as:continuity} and~\ref{as:karma-preservation} hold.
Then for every $\bar{k} \in \Nat$, there exists a \gls{SNE} $(d^*,\pi^*)$ satisfying
\begin{align*}
    \sum_{\tau \in \Gamma} \sum_{u \in \U_\tau} \sum_{k \in \Nat} d^*_\tau[u,k] \: k = \kbar,
\end{align*}
where $\kbar$ is the average amount of karma per agent in the system.
\end{theorem}

\section{Proofs}

\subsection{Proof of Proposition~\ref{prop:sne-existence}}
\label{proof:karma-preservation}
The proof is an application of Theorem~\ref{th:Nash-exists}, thus we must prove that Assumptions~\ref{as:continuity} and \ref{as:karma-preservation} are verified. 
It is straightforward to verify that, with the continuous approximation $\psi^\epsilon$, all the expressions in the derivation of $\zeta_\tau[u,t,b](d,\pi)$ and $\kappa[k^+ \mid k,t,b](d,\pi)$ are continuous in $(d,\pi)$.
Thus, Assumption~\ref{as:continuity} is satisfied.

Assumption~\ref{as:karma-preservation} is satisfied if \eqref{eq:karma-preservation} holds for the setup considered. Omitting the dependency on $(d,\pi)$, we have
\begin{align*}
    &\sum_{\tau \in \Gamma} \sum_{u \in \U_\tau} \sum_{k \in \Nat} d_\tau[u,k] \sum_{t \in \T} \sum_{b \in \B[k]} \pi_\tau[t,b \mid u,k] \sum_{k^+ \in \Nat} \kappa[k^+ \mid k,t,b] \: k^+ \\
    &\:\,\stackrel{\eqref{eq:bottleneck-karma-transitions},\eqref{eq:prob-outcome-cont}}{=}
    \sum_{\tau \in \Gamma} \sum_{u \in \U_\tau} \sum_{k \in \Nat} d_\tau[u,k] \sum_{t \in \T} \sum_{b \in \B[k]} \pi_\tau[t,b \mid u,k] \sum_{o \in \Out} \psi^\epsilon[o \mid t,b] \sum_{k^+ \in \Nat} \Prob[k^+ \mid k,t,b,o] \: k^+ \\
    &\quad
    \stackrel{\eqref{eq:bottleneck-karma-tranistions-o}}{=}
    \sum_{\tau \in \Gamma} \sum_{u \in \U_\tau} \sum_{k \in \Nat} d_\tau[u,k] \sum_{t \in \T} \sum_{b \in \B[k]} \pi_\tau[t,b \mid u,k] \big(\psi^\epsilon[o=\fast \mid t,b] \: (k - b + \pbar)  \\
    &\quad\phantom{= \sum_{\tau \in \Gamma} \sum_{u \in \U_\tau} \sum_{k \in \Nat} d_\tau[u,k] \sum_{t \in \T} \sum_{b \in \B[k]} \pi_\tau[t,b \mid u,k]}\quad  + \psi^\epsilon[o=\slow \mid t,b] \: (k + \pbar) \big)\\
    &\quad=
    \pbar + \sum_{\tau \in \Gamma} \sum_{u \in \U_\tau} \sum_{k \in \Nat} d_\tau[u,k] \: k - \sum_{\tau \in \Gamma} \sum_{u \in \U_\tau} \sum_{k \in \Nat} d_\tau[u,k] \sum_{t \in \T} \sum_{b \in \B[k]} \pi_\tau[t,b \mid u,k] \: \psi^\epsilon[o=\fast \mid t, b] \: b \\
    &\quad    \stackrel{\eqref{eq:average-payment}}{=}
    \sum_{\tau \in \Gamma} \sum_{u \in \U_\tau} \sum_{k \in \Nat} d_\tau[u,k] \: k.
\end{align*}
The equality from the second to the third line is due to {$(1-f)\floor{\pbar} + f\ceil{\pbar}=\pbar$. We conclude the proof by invoking Theorem~\ref{th:Nash-exists}.~\hfill\qed}

\subsection{Proof of Proposition~\ref{prop:equity}}
\label{proof:equity}

For ease of presentation, we consider two income types $\tau \in \Gamma = \{\low, \high\}$, and without loss of generality let $\lambda_\low=1$, $\lambda_\high = \lambda > 1$.
The proof holds for an arbitrary number of types using analogous arguments.

We prove the statement by constructing an equitable $\karma$ \gls{SNE}.
Consider an auxiliary income-homogeneous setting with the same total number of commuters where all are of the low income type, and let \mbox{$(\dhom,\pihom)$} be a corresponding $\karma$ \gls{SNE}, which is guaranteed to exist by Proposition~\ref{prop:sne-existence}.
Construct the social state $(\deq,\pieq)$ as
\begin{align*}
    &\deqtau = g_\tau \, \dhom, \; \textup{for all } \tau \in \Gamma, \\
    &\pieqlow = \pihom, \\
    &\pieqhigh[a \mid \lambda \, u, k] = \pihom[a \mid u,k], \; \textup{for all } [u,k,a] \in \Ulow \times \Nat \times \A[k].
\end{align*}
The remainder of the proof is devoted to showing that $(\deq,\pieq)$ is a \gls{SNE} in the heterogeneous setting, thus it satisfies~\eqref{eq:SNE-1}--\eqref{eq:SNE-2}.

\eqref{eq:SNE-1} is satisfied since $\dhom$ is stationary when the whole population follows $\pihom$, and thus $\deq$ is also stationary.

We therefore focus on showing that~\eqref{eq:SNE-2} holds, i.e., that $\pieq$ is optimal for both commuter types.
Observe that the elementary constituents of the commuter \gls{MDP}s, i.e., the immediate reward function $\zeta_\tau(d,\pi)$ and karma transition function $\kappa_\tau(d,\pi)$, depend on $(d,\pi)$ only through the bid distribution $\nu(d,\pi)$.
It is straight-forward to show that $\nu(\deq,\pieq) = \nu(\dhom,\pihom)$, therefore the low income commuters face an identical strategic problem under $(\deq,\pieq)$ and $(\dhom,\pihom)$ in which $\pieqlow$ is optimal.
For the high income commuters, it holds for all $[u,k,a,u^+,k^+] \in \Ulow \times \Nat \times \A[k] \times \Ulow \times \Nat$ at $(\deq,\pieq)$ (which we omit from the notation)
\begin{subequations}
\begin{align}
    \zeta_\high[\lambda \, u, a] &= \lambda \, \zeta_\low[u, a], \\
    R_\high[\lambda \, u, k] &= \lambda \, R_\low[u, k], \\
    \rho_\high[\lambda \, u^+, k^+ \mid \lambda \, u,k,a] &= \rho_\low[u^+, k^+ \mid u,k,a], \\
    P_\high[\lambda \, u^+, k^+ \mid \lambda \, u,k] & = P_\low[u^+, k^+ \mid u,k], \label{eq:proof-step-2} \\
    V_\high[\lambda \, u, k] &= \lambda \, V_\low[u, k], \label{eq:scaled-V} \\
    Q_\high[\lambda \, u, k, a] &= \lambda \, Q_\low[u, k, a]. \label{eq:scaled-Q}
\end{align}
\end{subequations}
Since $\pihom$ is optimal with respect to $Q_\low(\deq,\pieq)$, and $Q_\high(\deq,\pieq)$ differs from $Q_\low(\deq,\pieq)$ by the constant scaling factor $\lambda > 1$, it follows that $\pieqhigh$ is also optimal.
Finally, it holds that
\begin{align*}
    \cbarnorm_\high(\deq,\pieq) &= -\frac{1}{g_\high \, \ubar_\high} \: \sum_{u \in \U_\high} \sum_{k \in \Nat} \deqhigh[u,k] \: R_\high[u,k](\deq,\pieq) \\
    &= -\frac{1}{\lambda \, \ubar_\low} \: \sum_{u \in \U_\low} \sum_{k \in \Nat} \dhom[u,k] \: \lambda \, R_\low[u,k](\deq,\pieq) \\
    &= -\frac{1}{g_\low \, \ubar_\low} \: \sum_{u \in \U_\low} \sum_{k \in \Nat} \deqlow[u,k] \: R_\low[u,k](\deq,\pieq) \\
    &= \cbarnorm_\low(\deq,\pieq),
\end{align*}
which concludes the proof.
\hfill\qed

\subsection{Proof of Proposition~\ref{prop:equity-no-discount}}
\label{proof:equity-no-discount}

Analogously to the proof of Proposition~\ref{prop:equity}, without loss of generality we consider two income types $\tau \in \Gamma = \{\low, \high\}$, and let $\lambda_\low=1$, $\lambda_\high = \lambda > 1$.

Mutandi mutandis from~\cite[Proposition~4.1.2]{bertsekas2007dynamic}, the following holds for all $[\tau,u,k] \in \Gamma \times \U_\tau \times \Nat$,
\begin{multline}
    \cbarnorm_\tau[u,k](\ddelta,\pidelta) \\
    = -(1 - \delta) \, \frac{V_\tau[u,k](\ddelta,\pidelta)}{\ubar_\tau} + (1 - \delta) \, \frac{h_\tau[u,k](\ddelta,\pidelta)}{\ubar_\tau} + O((1 - \delta)^2),
    \label{eq:cbar-delta}
\end{multline}
where $\cbarnorm_\tau[u,k](\ddelta,\pidelta)$ is the normalized long-term average travel cost \emph{starting in state} $[u,k]$, and $h_\tau[u,k](\ddelta,\pidelta)$, defined as in~\cite[Eq.~4.16]{bertsekas2007dynamic}, is a so-called \emph{bias term} whose order of magnitude does not depend on $\delta$.
Thus, we have for all $[u,k] \in \U_\low \times \Nat$
\begin{align}
    &\lim_{\delta \rightarrow 1} \left[\cbarnorm_\high[\lambda \, u,k](\ddelta,\pidelta) - \cbarnorm_\low[u,k]\right(\ddelta,\pidelta)] \\
    &\quad= \lim_{\delta \rightarrow 1} \left[(1 - \delta) \left(\frac{h_\high[\lambda \, u,k]}{\ubar_\high} - \frac{h_\low[u,k]}{\ubar_\low}\right) + O((1 - \delta)^2)\right] \\
    &\quad = 0,
\end{align}
where we used~\eqref{eq:scaled-V} and the optimality of $\pideltahigh$ and $\pideltalow$.
This establishes the desired result conditioned that commuters of both income groups start from the same (normalized) state $[u,k] \in \U_\low \times \Nat$.

To conclude the proof we establish that for all $[\tau,u,k,u',k'] \in \Gamma \times \U_\tau \times \Nat \times \U_\tau \times \Nat$ it holds that
\begin{align}
    \lim_{\delta \rightarrow 1} \left[\cbarnorm_\tau[u,k](\ddelta,\pidelta) - \cbarnorm_\tau[u',k'](\ddelta,\pidelta)\right] = 0,
\end{align}
i.e., $\cbarnorm_\tau[u,k](\ddelta,\pidelta)$ does not depend on $[u,k]$ as $\delta \rightarrow 1$.
Using~\eqref{eq:cbar-delta} we have
\begin{multline}
     \lim_{\delta \rightarrow 1} \left[\cbarnorm_\tau[u,k](\ddelta,\pidelta) - \cbarnorm_\tau[u',k'](\ddelta,\pidelta)\right] \\
     = \lim_{\delta \rightarrow 1} \frac{(1 - \delta)}{\ubar_\tau} \left(V_\tau[u',k'](\ddelta,\pidelta) - V_\tau[u,k](\ddelta,\pidelta)\right).
\end{multline}
Then, from Lemma~\ref{lem:V-non-decreasing}, it follows that  for all $[\tau,u,k,k'] \in \Gamma \times \U_\tau \times \Nat \times \Nat$ it holds
\begin{align}
    \label{eq:cbar-non-increasing}
    \lim_{\delta \rightarrow 1} \left[\cbarnorm_\tau[u,k](\ddelta,\pidelta) - \cbarnorm_\tau[u,k'](\ddelta,\pidelta) \right] > 0 \Rightarrow k' > k.
\end{align}
Moreover, it is well known that the long-term average cost in a Markov chain is identical for all initial states belonging to the same communicating class, i.e., it differs per initial state only if the Markov chain has multiple communicating classes, see, e.g.,~\cite[Chapter~4.2]{bertsekas2007dynamic}.
Since the \gls{VOT} Markov chain $\phi_\tau$ is irreducible this is only possible if there are more than one communicating class in the karma states.
Suppose that this is the case, i.e., there exist $\tau \in \Gamma$, $u' \in \U_\tau$, $k',k'' \in \Nat$ such that
\begin{align*}
    \lim_{\delta \rightarrow 1} \left[\cbarnorm_\tau[u',k'](\ddelta,\pidelta) - \cbarnorm_\tau[u',k''](\ddelta,\pidelta)\right] > 0.
\end{align*}
Due to~\eqref{eq:cbar-non-increasing} we must have that $k'' > k'$, and similarly for all $k \geq k'' > k'$ we must have,
\begin{align*}
    \lim_{\delta \rightarrow 1} \left[\cbarnorm_\tau[u',k'](\ddelta,\pidelta) - \cbarnorm_\tau[u',k](\ddelta,\pidelta)\right] > 0.
\end{align*}
Thus, it must hold that $k'$ does not communicate with any $k \geq k''$.
Consider a candidate policy $\tilde \pi_\tau$ in which commuters bid 0 at all karma states $k\in [k' , k'' )$, and at karma states $k \geq k''$ follow policy $\pideltatau$ that yields $\lim_{\delta \rightarrow 1} \cbarnorm_\tau[u',k](\ddelta,\pidelta)$.
By Lemma~\ref{lem:positive-redistribution} this policy guarantees that some $k \geq k''$ is reachable from $k'$, after which a lower long-term average cost can be obtained.
Therefore, for $\delta \rightarrow 1$ this benefit will outweigh any transient costs incurred while reaching $k \geq k''$ from $k'$, hence $\tilde \pi_\tau$ will achieve a higher $V_\tau[u',k']$ than $\pideltatau$ that obtains $\lim_{\delta \rightarrow} \cbarnorm_\tau[u',k'](\ddelta,\pidelta)$.
This leads to a contradiction since $\pideltatau$ is an equilibrium policy. Thus, there is only one communicating class in the karma states and this concludes the proof.
\hfill\qed

\subsection{Supporting lemmas for Proposition~\ref{prop:strict-pareto}}
\label{sec:lemmas}

\begin{lemma}
\label{lem:V-non-decreasing}
At a $\karma$ SNE $(d^*,\pi^*)$, the value function is non-decreasing in karma, i.e., for all $[\tau, u, k] \in \Gamma \times \U \times \Nat$, 
\begin{align}
    V_\tau[u,k+1](d^*,\pi^*) \geq V_\tau[u,k](d^*,\pi^*).
\end{align}
\end{lemma}

\begin{proof}
By~\eqref{eq:SNE-2} $V_\tau(d^*,\pi^*)$ is an optimal value function, therefore it is the unique fixed point of the contraction mapping given for all $[\tau, u, k] \in \Gamma \times \U_\tau \times \Nat$ by, see, e.g., \cite[Chapter~1.4]{bertsekas2007dynamic},
\begin{align}
    &V_\tau^{(l+1)}[u,k] \notag \\
    &\quad= \max_{t \in \T, \, b \leq k} \left\{\zeta_\tau[u,t,b] + \delta \sum_{u^+ \in \U_\tau} \phi_\tau[u^+ \mid u] \sum_{k^+ \in \Nat} \kappa[k^+ \mid k,t,b] \, V_\tau^{(l)}[u^+,k^+]\right\} \notag \\ 
    &\quad= \max_{t \in \T, \, b \leq k} \left\{\zeta_\tau[u,t,b] + \delta \sum_{u^+ \in \U_\tau} \phi_\tau[u^+ \mid u] \sum_{\Delta k \in \Nat} \Prob[\Delta k \mid t,b] \, V_\tau^{(l)}[u^+,k + \Delta k]\right\}, \label{eq:value-iteration}
\end{align}
where we used $\Prob[\Delta k = k^+ - k \mid t, b]=\Prob[\Delta k = k^+ - k \mid k, t, b]:=\kappa[k^+ \mid k, t, b]$ and omitted $(d^*,\pi^*)$ from the notation.
Equation~\eqref{eq:value-iteration} is the well-known Bellman iteration.

We then proceed by induction over $l$.
Initialize with an arbitrary non-decreasing value function in karma, e.g., $V_\tau^{(0)}[u,k] = 0$ for all $[\tau, u, k] \in \T \times \U_\tau \times \Nat$.
As induction hypothesis assume that $V_\tau^{(l)}[u,k+1] \geq V_\tau^{(l)}[u,k]$ for all $[\tau, u, k] \in \T \times \U_\tau \times \Nat$.
Then it holds for $l+1$,
\begin{align*}
    &V_\tau^{(l+1)}[u,k+1] \\
    &\quad= \max_{t \in \T, \, b \leq k+1} \left\{\zeta_\tau[u,t,b] + \delta \sum_{u^+ \in \U_\tau} \phi_\tau[u^+ \mid u] \sum_{\Delta k \in \Nat} \Prob[\Delta k \mid t,b] \, V_\tau^{(l)}[u^+,k + 1 + \Delta k]\right\} \\
    &\quad \geq \max_{t \in \T, \, b \leq k} \left\{\zeta_\tau[u,t,b] + \delta \sum_{u^+ \in \U_\tau} \phi_\tau[u^+ \mid u] \sum_{\Delta k \in \Nat} \Prob[\Delta k \mid t,b] \, V_\tau^{(l)}[u^+,k + 1 + \Delta k]\right\} \\
    &\quad \geq \max_{t \in \T, \, b \leq k} \left\{\zeta_\tau[u,t,b] + \delta \sum_{u^+ \in \U_\tau} \phi_\tau[u^+ \mid u] \sum_{\Delta k \in \Nat} \Prob[\Delta k \mid t,b] \, V_\tau^{(l)}[u^+,k + \Delta k]\right\} \\
    &\quad = V_\tau^{(l+1)}[u,k],
\end{align*}
as needed.
Letting $l \rightarrow \infty$ establishes the result.
\end{proof}

\begin{lemma}
\label{lem:positive-redistribution}
At a $\karma$ SNE $(d^*,\pi^*)$, the total karma redistributed to commuters is strictly positive. Due to the uniform redistribution scheme, this is equivalent to a positive average payment $\pbar(d^*,\pi^*) > 0$.
\end{lemma}

\begin{proof}
We omit $(d^*,\pi^*)$ for notational convenience.
Suppose, for the sake of contradiction, that $\pbar = 0$.
Then it must hold that $\pi^*_\tau[t,b>0 \mid u,k]=0$ for all $[\tau,u,k,t] \in \Gamma \times \U_\tau \times \Nat \times \T$ with $d^*_\tau[u,k] > 0$.
Moreover, since with $\pbar = 0$ there are no karma dynamics and the \gls{VOT} dynamics are exogenous, using a similar argument as in the proof of Proposition~\ref{prop:strict-pareto} it is straightforward to show that the value function $V_\tau[u,k]$ depends on $u$ only and maximizing the state-action value function $Q_\tau[u,k,a]$ is equivalent to maximizing the immediate rewards $\zeta_\tau[u,a]$.
It must also hold that $q[t^*] > 0$; otherwise action $[t^*,b=0]$ incurs zero immediate cost and strongly dominates all other actions leading to $q[t^*] > 0$.

Since all commuters are initially endowed with a positive amount of karma, there must exist $[\tilde \tau, \tilde u, \tilde k \geq 1]$ for which $d^*_{\tilde \tau}[\tilde u, \tilde k] > 0$.
But action $[t^*,b=1]$ guarantees zero immediate cost and strongly dominates all actions with $b=0$, contradicting that $\pi^*_{\tilde \tau}$ is an equilibrium policy and concluding the proof.
\end{proof}

\begin{lemma}
\label{lem:weak-Pareto}
Under Assumptions~\ref{as:same-t*} and \ref{as:discretization}, there is no queue at $t=\tstart$ at any $\karma$ \gls{SNE} $(d^*,\pi^*)$. 
As a consequence, the expected immediate rewards satisfy for all $[\tau,u,k] \in \Gamma \times \U_\tau \times \Nat$,
\begin{align}
    \label{eq:weak-Pareto}
    R_\tau[u,k](d^*,\pi^*) \geq -u \: c^*.
\end{align}
\end{lemma}

\begin{proof}
We omit $(d^*,\pi^*)$ for notational convenience and first show that there is no queue at $t=\tstart$.
Suppose, for the sake of contradiction, that $q[\tstart] > 0$.
Then the following conditions must hold:
\begin{enumerate}
    \item There exists $[\tilde \tau, \tilde u, \tilde k, \tilde b] \in \Gamma \times \U_{\tilde \tau} \times \Nat \times \B[\tilde k]$ such that $\pi^*_{\tilde \tau}[\tstart,\tilde b \mid \tilde u, \tilde k] > 0$ and
    \begin{subequations}
    \label{eq:weak-pareto-proof}
    \begin{align}
        &Q_{\tilde \tau}[\tilde u, \tilde k, \tstart, \tilde b] \notag\\
        &\quad= -\zeta_{\tilde \tau}[\tilde u, \tstart, \tilde b] + \delta \sum_{u^+ \in \U_{\tilde \tau}} \phi_{\tilde \tau}[u^+ \mid \tilde u] \sum_{k^+ \in \Nat} \kappa[k^+ \mid \tilde k, \tstart, \tilde b] \: V_{\tilde \tau}[u^+,k^+] \notag \\
        &\quad< -\tilde u \: c^* + \delta \sum_{u^+ \in \U_{\tilde \tau}} \phi_{\tilde \tau}[u^+ \mid \tilde u] \sum_{k^+ \in \Nat} \kappa[k^+ \mid \tilde k, \tstart, \tilde b] \: V_{\tilde \tau}[u^+,k^+] \label{eq:weak-pareto-proof-1} \\
        &\quad\leq -\tilde u \: c^* + \delta \sum_{u^+ \in \U_{\tilde \tau}} \phi_{\tilde \tau}[u^+ \mid \tilde u] \: V_{\tilde \tau}[u^+, \tilde k + \pbar], \label{eq:weak-pareto-proof-2}
    \end{align}
    \end{subequations}
    where we defined $V_{\tilde \tau}[u^+, \tilde k + \bar p] := f \: V_{\tilde \tau}[u^+, \tilde k + \ceil{\bar p}] + (1 - f) \: V_{\tilde \tau}[u^+, \tilde k + \floor{\bar p}]$.
    Inequality~\eqref{eq:weak-pareto-proof-1} holds because $c^* = \beta \: (t^* - \tstart)$ is the queue-free normalized cost at $\tstart$ and therefore a strictly greater immediate cost is incurred when $q[\tstart] > 0$, and~\eqref{eq:weak-pareto-proof-2} holds due to Lemma~\ref{lem:V-non-decreasing} and that $\tilde k + \pbar$ is the maximum possible next karma.

    \item For all $t \in \{\tstart + \Delta,\dots,\tend\}$, $q[t] > 0$.
    Otherwise, if $q[t'] = 0$ for some $t' \in \{\tstart + \Delta,\dots,\tend\}$, it is straight-forward to verify that $Q_{\tilde \tau}[\tilde u, \tilde k, t', b=0]$ is greater or equal to~\eqref{eq:weak-pareto-proof-2}, contradicting that $[\tstart,\tilde b]$ is an equilibrium action.

    \item Similarly, for all $t \in \{\tstart,\dots,\tend\}$, the fast lane is fully occupied, i.e., $N^\fast[t] = \sfast \: \Delta$, where $N^\fast[t]$ is the number of commuters entering the fast lane at time $t$\footnote{With minor loss of generality it is assumed that the continuous approximation parameter $\epsilon$ is small enough to have a negligible effect in under-allocating the fast lane. Formally one can write $N^\fast[t] = \sfast \: \Delta - O(\epsilon)$, which does not affect the result for $\epsilon \rightarrow 0$.}.
    Otherwise, if it is not fully occupied for some $t' \in \{\tstart,\dots,\tend\}$, it holds that $b^*[t'] = 0$ and action $[t',b=0]$ guarantees entry into the fast lane, achieving $Q_{\tilde \tau}[\tilde u, \tilde k, t', b=0]$ greater or equal to~\eqref{eq:weak-pareto-proof-2} and contradicting that $[\tstart,\tilde b]$ is an equilibrium action.

    \item Let $\hat t = \max \{t < \tstart \mid q[t] = 0\}$ be the latest queue-free departure time preceding $\tstart$. For all $t \in \{\hat t + \Delta,\dots,\tend\}$, since $q[t] = \max\{0, q[t-\Delta] + n^\slow[t] - \sslow \: \Delta\} > 0$, we have $N^\slow[t] = \sslow \: \Delta + q[t] - q[t-\Delta]$, where $N^\slow[t]$ is the number of commuters entering the slow lane at time $t$.
\end{enumerate}
It follows that the total number of departures in $t \in \{\hat t + \Delta, \dots,\tend\}$ is
\begin{subequations}
    \begin{align*}
        &\sum_{t=\hat t + \Delta}^{\tend} \left(N^\slow[t] + N^\fast[t]\right) \\
        &\quad\geq \sum_{t=\hat t + \Delta}^{\tend} N^\slow[t] + \sum_{t=\tstart}^{\tend} N^\fast[t] \\
        &\quad= \sum_{t=\hat t + \Delta}^{\tend} \left(\sslow \: \Delta + q[t] - q[t-\Delta]\right) + (\tend - \tstart + \Delta) \: \sfast \\
        &\quad= (\tend - \hat t) \: \sslow + q[\tend] + (\tend - \tstart + \Delta) \: \sfast \\
        &\quad\geq (\tend - \tstart + \Delta) \: (\sslow + \sfast) + q[\tend] \\
        &\quad= N + s \: \Delta + q[\tend] > N,
        \end{align*}
\end{subequations}
leading to a contradiction. Thus it must hold that $q[\tstart] = 0$.

Next we show~\eqref{eq:weak-Pareto}.
Suppose, for the sake of contradiction, that there exists some $[\tilde \tau,\tilde u,\tilde k]$ for which $-R_{\tilde \tau}[\tilde u, \tilde k] > \tilde u \: c^*$.
Then there must exist action $\tilde a$ with $\pi^*_{\tilde \tau}[\tilde a \mid \tilde u, \tilde k] > 0$ for which $\zeta_{\tilde \tau}[\tilde u, \tilde a] < -\tilde u \: c^*$, and
\begin{align}
    \label{eq:weak-pareto-proof-end}
    Q_{\tilde \tau}[\tilde u, \tilde k, \tilde a] &< -\tilde u \: c^* + \delta \sum_{u^+ \in \U_{\tilde \tau}} \phi_{\tilde \tau}[u^+ \mid \tilde u] \: V_{\tilde \tau}[u^+, \tilde k + \bar p],
\end{align}
following similar arguments as in~\eqref{eq:weak-pareto-proof}.
But since $q[\tstart]=0$, action $[\tstart,b=0]$ achieves $Q_{\tilde \tau}[\tilde u,\tilde k,\tstart,b=0]$ equal to the right-hand side of~\eqref{eq:weak-pareto-proof-end}.
This contradicts that $\tilde a$ is an equilibrium action and concludes the proof.
\end{proof}

\subsection{Proof of Proposition~\ref{prop:strict-pareto}}
\label{proof:strict-pareto}

Lemma~\ref{lem:weak-Pareto} directly implies that a weak version of Proposition~\ref{prop:strict-pareto}  holds, i.e., $\cbarnorm_\tau(d^*,\pi^*) \leq c^*$ for all $\tau \in \Gamma$.
We complete the proof by showing that this inequality can be made strict.
We omit $(d^*,\pi^*)$ for notational convenience.
For an arbitrary $\tau \in \Gamma$ suppose, for the sake of contradiction, that $\cbarnorm_\tau = c^*$. Then, due to Lemma~\ref{lem:weak-Pareto} it must hold that $R_\tau[u,k] = -u \: c^*$ for all $[u,k] \in \U_\tau \times \Nat$ satisfying $d^*_\tau[u,k] > 0$, i.e., $R_\tau[u,k]$ depends on $u$ only and not on $k$, which we denote by $R_\tau[u,k] = R_\tau[u]$.
We show that $V_\tau[u,k] = V_\tau[u]$ must also depend on $u$ only and not on $k$.
It is well known that $V_\tau[u,k]$ is the unique fixed point of the contraction mapping given for all $[u, k] \in \U_\tau \times \Nat$ by, see, e.g.,~\cite[Chapter~1.4]{bertsekas2007dynamic},
\begin{align}
    \label{eq:value-iteration-fixed}
    V_\tau^{(l+1)}[u,k] = R_\tau[u] + \delta \sum_{u^+ \in \U_\tau} \phi_\tau[u^+ \mid u] \sum_{a \in \A[k]} \pi_\tau[a \mid u,k] \sum_{k^+ \in \Nat} \kappa[k^+ \mid k, a] \: V_\tau^{(l)}[u^+,k^+].
\end{align}
Then, we show that $V_\tau[u,k] = V_\tau[u]$ by induction over $l \in \Nat$. Initialize with any $V_\tau^{(0)}[u,k] = V_\tau^{(0)}[k]$, e.g., $V_\tau^{(0)}[u, k] = 0$ for all $[u,k] \in \U_\tau \times \Nat$.
As induction hypothesis assume that $V_\tau^{(l)}[u,k] = V_\tau^{(l)}[u]$.
Then it holds for $l+1$,
\begin{align*}
    V_\tau^{(l+1)}[u,k] &= R_\tau[u] + \delta \sum_{u^+ \in \U_\tau} \phi_\tau[u^+ \mid u] \sum_{a \in \A[k]} \pi_\tau[a \mid u,k] \sum_{k^+ \in \Nat} \kappa[k^+ \mid k, a] \: V_\tau^{(l)}[u^+] \\
    &= R_\tau[u] + \delta \sum_{u^+ \in \U_\tau} \phi_\tau[u^+ \mid u] \: V_\tau^{(l)}[u^+] = V_\tau^{(l+1)}[u],
\end{align*}
as needed.

Similarly, for the state-action value function we have
\begin{align*}
    Q_\tau[u,k,a] &= \zeta_\tau[u,a] + \delta \sum_{u^+ \in \U_\tau} \phi_\tau[u^+ \mid u] \sum_{k^+ \in \Nat} \kappa[k^+ \mid k, a] \: V_\tau[u^+] \\
    &= \zeta_\tau[u,a] + \delta \sum_{u^+ \in \U_\tau} \phi_\tau[u^+ \mid u] \: V_\tau[u^+].
\end{align*}
Notice that, since the \gls{VOT} process $\phi_\tau[u^+ \mid u]$ is exogeneous, maximizing $Q_\tau[u,k,a]$ with respect to the action $a$ is equivalent to maximizing the immediate rewards $\zeta_\tau[u,a]$.

Furthermore, it must hold that the type $\tau$ commuters never enter the fast lane for sure during a time $t \in \T^\textup{ben} := \{\tstart+\Delta,\dots,\tend-\Delta\}$; otherwise they would experience an immediate reward that is strictly greater than $-u \: c^*$, since $c^*$ equals the normalized queue-free cost at $\tstart$ or $\tend$.
Simultaneously, due to Lemma~\ref{lem:positive-redistribution} a type $\tau$ commuter has a non-zero probability of gaining a positive amount of karma every day.
Therefore, for any $\tilde t \in \T^\textup{ben}$, there is a non-zero probability of eventually reaching a state $[\tilde u, \tilde k]$, where $\tilde k \geq b^*[\tilde t] + 1$, i.e., it holds that $d^*_\tau[\tilde u, \tilde k] > 0$. At such a state, an action $[\tilde t,b^*[\tilde t] + 1]$ guarantees entry into the fast lane, and an immediate reward that is strictly greater than $-\tilde u \: c^*$, which leads to a contradiction and concludes the proof.
\hfill\qed

\subsection{Proof of Proposition~\ref{prop:unique-redistribution}}
\label{proof:unique-redistribution}

Given $n[t](d,\pi)$ for all $t \in \T$ and $\pbar(d,\pi)$, \eqref{eq:time-redistribution}--\eqref{eq:time-redistribution-karma-preservation} constitute a linear system of $T$ equations in the $T$ unknowns $r[t](d,\pi)$ for all $t \in \T$.
In vector form, this can be written as
\begin{align*}
\Are(d,\pi) \: r(d,\pi) = \bre(d,\pi),
\end{align*}
where, omitting $(d,\pi)$ from the notation,
\begin{align*}
\Are &= \begin{bmatrix}
    1 & \cdots & 0 & -m[t^1] & 0 & \cdots &  0 \\
    \vdots & \ddots & \vdots & \vdots & \vdots & \ddots & \vdots \\
    0 & \cdots & 1 & -m[\tminre - \Delta] & 0 & \cdots & 0 \\[5mm]
    0 & \cdots & 0 & -m[\tminre + \Delta] & 1 & \cdots & 0 \\
    \vdots & \ddots & \vdots & \vdots & \vdots & \ddots & \vdots \\
    0 & \cdots & 0 & -m[t^T] & 0 & \cdots & 1 \\[3mm]
    n[1] & \cdots & n[\tminre - \Delta] & n[\tminre] & n[\tminre + \Delta] & \cdots & n[T]
\end{bmatrix},\\
\bre &= \begin{bmatrix}
    0 \\ \vdots \\ 0 \\ \pbar
\end{bmatrix}.
\end{align*}

The first $T-1$ rows of $\Are(d,\pi)$ are linearly independent.
To verify the linear independence of the last row $T$, we proceed with Gaussian elimination.
In the reduced matrix, row $T$ has all zeros except in the $\tminre$-th position, which equals
\begin{align}
\label{eq:redistribution-matrix-proof}
\sum_{t\in \mc T} m[t] \: n[t](d,\pi) > 0,
\end{align}
where we used $m[\tminre]=1$.
Inequality~\eqref{eq:redistribution-matrix-proof} holds because $m[t] > 0$ and $n[t](d,\pi) \geq 0$ for all $t \in \T$, and for each $(d,\pi) \in \D \times \Pi$ there exists at least one $\tilde t \in \T$ for which $n[\tilde t](d,\pi) > 0$ (since $\sum_{t \in \T} n[t](d,\pi) = 1$).
This implies that $\Are(d,\pi)$ is invertible. Accordingly, the redistribution vector is unique and reads as 
\begin{align*}
r(d,\pi) = \Are^{-1}(d,\pi) \: \bre(d,\pi).
\end{align*}
It is also continuous in $(d,\pi)$ as $\Are^{-1}(d,\pi)$ and $\bre(d,\pi)$ are continuous in $(d,\pi)$.
\hfill\qed

 \bibliographystyle{elsarticle-num} 
 \bibliography{main}




\end{document}